\documentclass[aip, pof]{revtex4-2}
\usepackage{amssymb,amsmath,amsthm,amsfonts}
\usepackage{subcaption}
\usepackage{bm}
\usepackage{graphicx}
\usepackage{xcolor}

\begin{document}

\title{A Novel Velocity Discretization for Lattice Boltzmann Method: Application to Compressible Flow} 



\author{Navid Afrasiabian}

\author{Colin Denniston}
\affiliation{Department of Physics and Astronomy, University of Western Ontario, London, Ontario, Canada, N6A 3K7}

\date{\today}

\begin{abstract}
The Lattice Boltzmann Method (LBM) has emerged as a powerful tool in computational fluid dynamics and material science. However, standard LBM formulation imposes some limitations on the applications of the method, particularly compressible fluids. In this paper, we introduce a new velocity discretization method to overcome some of these challenges. In this new formulation, the particle populations are discretized using a bump function that has a mean and a variance. This introduces enough independent degrees of freedom to set the equilibrium moments to the moments of Maxwell-Boltzmann distribution up to and including the third moments. Consequently, the correct macroscopic fluid dynamics equations for compressible fluids are recovered. We validate our method using several benchmark simulations.
\end{abstract}

\maketitle 

\section{Introduction}\label{sec:intro}
The Boltzmann and Bhatnagar–Gross–Krook (BGK) equations are analytically applied to various systems such as dilute gases (including mixtures), dilute Lorentz gases, dense hard-sphere systems, Boltzmann-Langevin systems (which naturally incorporate stochastic fluctuations), and granular gases \cite{Dorfman2021book}. These systems are typically compressible fluids where energy flow arises naturally.

In contrast, the Lattice Boltzmann (LB) method, the most widely used application of the Boltzmann equation to fluid dynamics, is predominantly employed in the incompressible limit and for complex fluids, rather than for phenomena in compressible fluids like sound waves in gases. This limitation stems from certain issues in the standard LB method's formulation which we describe below.

The linearized Boltzmann equation (i.e. the BGK approximation of the Boltzmann equation) describes the evolution of the particle distribution function $f(\bm{x}, \bm{v}, t)$ towards the equilibrium state $f^{eq}(\rho, \bm{u}, T, \bm{v})$
\begin{equation}\label{eq:LinBoltz}
    \partial_t f + v_\alpha \partial_\alpha f = -\frac{1}{\tau} (f-f^{eq})
\end{equation}
where $\bm{v}$ is the microscopic velocity, $\bm{x}$ is position in space, and $t$ is time. The equilibrium particle distribution has a Maxwell-Boltzmann (MB) form
\begin{equation}
    f^{eq} = \frac{\rho}{(2 \pi T)^{d/2}}\text{exp}\left(-\frac{|\bm{v}-\bm{u}|^2}{2T}\right)
\end{equation}
where $\rho(\bm{x}, t)$ and $\bm{u}(\bm{x}, t)$ are the local equilibrium fluid density and mean velocity, respectively. $T$ is the temperature (in energy units) and $d$ is the dimensionality. In the standard Lattice Boltzmann method, the Hermite polynomials are used to discretize the particle distribution function as $f^{eq}$ has the same form as the generating function of the Hermite polynomials. A truncated local equilibrium distribution expanded around the global equilibrium is
\begin{equation}\label{eq:truncated_MB}
    f^{eq} \approx \rho g(\bm{v}) \left[1+v_\alpha u_\alpha + (u_\alpha u_\beta +(T-1)\delta_{\alpha \beta})(v_\alpha v_\beta -\delta_{\alpha \beta})\right]
\end{equation}
where $g(\bm{v})$ is the Hermite generating function. Using Gauss-Hermite quadrature, one can recover the exact equilibrium moments up to a certain degree depending on the discrete velocity set and number of discrete velocities used. The discrete equilibrium distribution is then
\begin{equation}\label{eq:old_discrete_f}
    f_i^{eq} = \rho w_i \left[1+\frac{e_{i\alpha} u_\alpha}{c^2} + \frac{1}{2c^4}(u_\alpha u_\beta +(T-1)\delta_{\alpha \beta})(e_{i\alpha} e_{i\beta} -\delta_{\alpha \beta})\right]
\end{equation}
where $\bm{e}_i$ is the discrete velocity by which the discrete distribution $f_i$ travels to neighbouring sites and $c=\Delta x / \Delta t$ is the lattice velocity. The challenge of capturing the correct hydrodynamics lies in reproducing the required number of moments using a finite number of discrete velocities.

The Chapman-Enskog method is a perturbation method that provides a framework to connect Boltzmann equation to the equations of hydrodynamics \cite{Dorfman2021book}. If we rearrange the Boltzmann equation to write $f$ in terms of $f^{eq}$ and its derivatives, we get
\begin{equation}
    f = f^{eq} -\tau D f
\end{equation}
where $D := \partial_t + v_\alpha \partial_\alpha $ is the material derivative. If we expand this term by recursively plugging the definition of $f$, we get
\begin{align}
    f &= f^{eq} - \tau D f^{eq} + \tau^2 D^2 f\\
    f &= f^{eq} - \tau D f^{eq} + \tau^2 D^2 f^{eq} - \tau^3 D^3 f
\end{align}
and so on. It has been shown that to recover the Navier-Stokes equations, one needs only up to second order derivatives of $f^{eq}$ \cite{Kruger2016book}. Thus,
\begin{equation}
    f \approx f^{eq} - \tau D f^{eq} + \tau^2 D^2 f^{eq}.
\end{equation}
This implies that to recover the continuity equation, we need the correct moments up to the second moment. Similarly, we would need up to reproduce the correct moments up to the third moment to recover the correct viscous stress terms. However, the discretization described by Eq. (\ref{eq:old_discrete_f}) is not capable of this for $e_{i\alpha}$ going out to the nearest neighbours (i.e. $e_{i\alpha} \in \{0,c\}$ for $\alpha = x, y, z$). This is because on standard lattices (nearest neighbour lattices), the higher order moments are dependent on the lower order moments. For example, the third diagonal moments of $f_i^{eq}$ are written as
\begin{equation}
    \sum_i f_i^{eq} e_{i\alpha} e_{i\beta} e_{i\gamma}|_{\alpha=\beta=\gamma} = \sum_i f_i^{eq} e_{i\alpha} e_{i\alpha} e_{i\alpha} = c^2 \sum_i f_i^{eq} e_{i\alpha} 
\end{equation}
since $e_{i\alpha} = 0 \text{ or } c$ (no sum over $\alpha$ is assumed here). To retain isotropic properties required by the hydrodynamic equations, one then has to enforce this condition on all third moments (even the ones that technically could be independently set) leading to 
\begin{equation}
    \sum_i f_i^{eq} e_{i\alpha} e_{i\beta} e_{i\gamma} = \frac{\rho}{3} c^2 (u_\alpha \delta_{\beta \gamma} + u_\gamma \delta_{\alpha \beta}+ u_\beta \delta_{\gamma \alpha})
\end{equation}
In this case the momentum equation takes the form of
\begin{align}\label{eq:standard_momentum}
    \partial_t (\rho u_\beta) + \partial_\alpha (\rho u_\alpha u_\beta) = - \partial_\beta P_{\alpha \beta} &+ \partial_\alpha \left(\rho \tau \frac{c^2}{3} (\partial_\beta u_\alpha +\partial_\alpha u_\beta + \partial_\gamma u_\gamma \delta_{\alpha \beta})\right)\nonumber\\
    &+ \partial_\alpha \left(\frac{c^2}{3} \tau (u_\beta \partial_\alpha \rho + u_\alpha \partial_\beta \rho + u_\gamma \delta_{\alpha \beta} \partial_\gamma \rho)\right)\nonumber\\
    &- \tau \partial_\alpha \left(u_\beta \partial_\gamma P_{\gamma \alpha} + u_\alpha \partial_\gamma P_{\gamma \beta} +\partial_\gamma (\rho u_\alpha u_\beta u_\gamma)\right)\nonumber\\
    &- \tau \partial_\alpha \left(\partial_\rho P_{\alpha \beta} \partial_\gamma(\rho u_\gamma) \right).
\end{align}
The first line in Eq. (\ref{eq:standard_momentum}) is the Navier-Stokes equation and all the other terms are error terms due to standard discretization of the Boltzmann equation on a standard lattice. These error terms also introduce violations of Galilean invariance. Some of these error terms can be canceled out if the equation of state $P_{\alpha \beta} = \rho \frac{c^2}{3} \delta_{\alpha \beta}$ is used. However, this introduces a coupling between the equation of state and the lattice velocity and would not solve all the problems as the cubic term $-\partial_\gamma (\rho u_\alpha u_\beta u_\gamma)$ would still be present.

Some techniques have been proposed to minimize or eliminate the impact of the error terms \cite{Sun1998a, Sun2000,Guangwu1999,Kataoka2004}. Alexander et al. proposed an LBM where the equilibrium distributions were adjusted so that a flexible speed of sound can be set \cite{Alexander1992}. As they pointed out, their model was limited to isothermal compressible flows.

Ji et al. proposed a Finite Volume-Lattice Boltzmann method for compressible flows \cite{Ji2009}. FVM is one of the most popular methods in computational fluid dynamics. FVM is designed based on conservation of quantities inside a cell in the system. Therefore, any changes in a quantity is balanced by that quantity's flux. This requires that the fluxes are evaluated at the cell interface. There are different methods to evaluate the flux. The proposed method by Ji et al. used LB to solve the local Riemann problem across the interface. Their model showed superiority compared to the conventional Godunov scheme and was able to capture shock wave, contact discontinuity and rarefaction waves.

Karlin et al. introduced the consistent LB model for weakly compressible flows. \cite{Ansumali2005}. In this model, they used an $\mathcal{H}$-theorem and applied the Gauss-Hermite quadrature to the $\mathcal{H}$-function. By including the conservation of energy, they were able to remove the spurious bulk viscosity present in previous isothermal models. Also, they were able to achieve this on standard lattices which made their model more efficient compared to multispeed models.

To address the lack of Galilean invariance in conventional LBM, Frapolli et al. proposed the co-moving Galilean Reference Frame \cite{Frapolli2016}. They argued that the errors and numerical instabilities observed in conventional LBM for flows at high Mach numbers are due to streaming particle populations with fixed discrete velocities, i.e. using the ``at rest" reference frame. The co-moving Galilean Reference Frame introduces uniformly shifted lattices, where the reference frame is consistently shifted across the numerical domain. They found that this approach works well for unidirectional compressible flows but is less effective for flows with significant variations in velocity and temperature \cite{Kallikounis2024}.

Dorschner et al. suggested that these limitations can be removed if tailored discrete velocities rather than fixed velocities were used. Based on this principle, they introduced the ``Particle on Demand" or the PonD model \cite{Dorschner2018}. In the PonD model, the kinetic equations are reconstructed in local reference frames defined by the actual local fluid velocity and temperature. In the streaming step, a predictor-corrector iteration loop is implemented to find the particle populations in the co-moving reference frame. The co-moving reference frame is the reference frame/gauge in which the particle velocities are determined by the fluid velocity and temperature at the monitored site. Their model was able to capture Galilean invariance and conserve mass, momentum, and energy.

In this paper, we introduce a new interpretation of the standard velocity discretization of LBM and propose a new discretization method based on this new interpretation. The new velocity discretization introduces extra independent degrees of freedom to recover the second and third moments of the Maxwell-Boltzmann distribution fully eliminating the error terms in Eq. (\ref{eq:standard_momentum}). Several standard examples of compressible fluids are then used to validate the model.

The paper is organized as follows: A new interpretation of the standard LBM velocity discretization in terms of delta functions is provided in section \ref{subsec:delta}. We describe our new discretization method based on this new interpretation in section \ref{subsec:bs}. The space and time discretization for the new method is discussed in section \ref{subsec:time_discretization} followed by a Chapman-Enskog analysis that shows that our model recovers the Navier-Stokes equations without the error terms found in the standard models. In section \ref{sec:application}, we apply our LBM to several benchmark examples such as Poiseuille flow, sound wave decay, Couette flow in a gravitational field, and flow over a cylinder. We finally summarize and conclude the paper in section \ref{sec:conclusion}.

\section{Theory}\label{sec:theory}
\subsection{New Interpretation on Standard Velocity Discretization} \label{subsec:delta}
We start with the ansatz that the particle distribution function $f$ can be expanded in terms of delta functions
\begin{equation}
    f(\bm{x}, \bm{v}, t) = \sum_i f_i \delta (\bm{v}-\bm{e}_i)
    \label{deltaf}
\end{equation}
where $f_i = f(\bm{e}_i)$ is the probability mass function. If we plug the expansion into the Linearized Boltzmann equation in Eq. (\ref{eq:LinBoltz}) (i.e. we will assume that this continuous equation is the foundation of the method rather than the discrete LB eqn) and integrate over the velocity space $\Omega_v$, the left-hand side (LHS) takes the form
\begin{align}
    \int_{\Omega_v} \left(\partial_t f + v_\alpha \partial_\alpha f\right) d\bm{v} &= \int_{\Omega_v} \left(\partial_t \sum_i f_i \delta (v_\alpha-e_{i\alpha}) + v_\alpha \partial_\alpha \sum_i f_i \delta (v_\alpha-e_{i\alpha}) \right) d\bm{v} \nonumber\\
    &=  \sum_i \left(\partial_t \int_{\Omega_v}  f_i \delta (v_\alpha-e_{i\alpha}) d\bm{v} + \int_{\Omega_v} v_\alpha \partial_\alpha  f_i \delta (v_\alpha-e_{i\alpha}) d\bm{v} \right)\nonumber \\
    &=  \sum_i \left(\partial_t  f_i + e_{i\alpha} \partial_\alpha  f_i \right)
\end{align}
and the right-hand side (RHS) becomes
\begin{equation}
   \int_{\Omega_v} -\frac{1}{\tau} (f-f^{eq}) d\bm{v} = -\frac{1}{\tau} \int_{\Omega_v} \sum_i (f_i-f_i^{eq}) \delta (v_\alpha-e_{i\alpha})= \sum_i -\frac{1}{\tau} (f_i-f_i^{eq}).
\end{equation}
We use a single-relaxation time model in this paper for sake of simplicity, but the extension to multi-relaxation time model should be straightforward. Putting this together, we get an equation that describes the evolution of the lowest moment of $f$. 
\begin{equation}
	\sum_i \left\{ \left(\partial_t f_i + e_{i\alpha} \partial_\alpha  f_i \right)=-\frac{1}{\tau} (f_i-f_i^{eq})\right\}
\end{equation}
 If we insist that for every $i$ the above equations holds (stricter condition than the sum), we obtain the discrete velocity Boltzmann equation:
\begin{equation}
    \partial_t  f_i + e_{i\alpha} \partial_\alpha  f_i = -\frac{1}{\tau} (f_i-f_i^{eq}),\text{ for } \forall \ i
    \label{stdlb}
\end{equation}
The  evolution of higher moments can similarly be obtained by substituting Eq. (\ref{deltaf}) into the continuous Linearized Boltzmann equation Eq. (\ref{eq:LinBoltz}), multiplying by powers of $v_\beta$ ({\it not} $e_{i\beta}$), and then integrating over $v$ to get :
\begin{align}
    &\sum_i \left\{\partial_t  (f_i e_{i\beta}) + e_{i\alpha} \partial_\alpha  (f_i e_{i\beta}) = -\frac{1}{\tau} (f_i e_{i\beta}-f_i^{eq} e_{i\beta})\right\},\\
    &\sum_i \left\{\partial_t  (f_i e_{i\beta} e_{i\gamma})+ e_{i\alpha} \partial_\alpha  (f_i e_{i\beta} e_{i\gamma}) = -\frac{1}{\tau} (f_i e_{i\beta}e_{i\gamma}-f_i^{eq} e_{i\beta}e_{i\gamma}) \right\}.
\end{align}
Again, these are the equations used in the standard LBM. Therefore, we can see that the standard model is equivalent to using an expansion of the particle distribution function $f$ in terms of delta functions. Note that in this case the evolution equations of the $n$th moment for the discrete velocity LB can be recovered by simply multiplying the discrete velocity LB Eq. (\ref{stdlb}) by $e_i$'s and summing.  As a result, one can chose to just employ Eq.(\ref{stdlb}) as written or, as is commonly done, carry out the collision term in moment space. The moments of $f$ are 
\begin{align}
    &\int f d\bm{v} = \sum_i f_i \\
    &\int f v_\alpha d\bm{v} = \sum_i f_i e_{i\alpha}\\
    &\int f v_\alpha v'_\beta d\bm{v} d\bm{v}' = \sum_i f_i e_{i\alpha} e_{i\beta}\\
    &\int f v_\alpha v'_\beta v''_\gamma d\bm{v} d\bm{v}' d\bm{v}'' = \sum_i f_i e_{i\alpha} e_{i\beta} e_{i\gamma}.
\end{align}
This is where the challenge of the standard LBM presents itself. Ideally, we would like to set the moments of the equilibrium distribution function $f_i^{eq}$ to the moments of the Maxwell-Boltzmann distribution. However, as mentioned in the introduction, the above formulation is not capable of matching the moments of the MB distribution beyond the second moment on standard lattices (i.e. nearest neighbours only).

The new interpretation provided by attempting to solve the continuous linearized Boltmann equation using an expansion of delta function distributions suggests a potential solution that is: What if we did not use a delta function but a different probability density function to estimate our $f$? In the next section, we describe and explain this new method.

\subsection{New Velocity Discretization}\label{subsec:bs}
We saw that an expansion of the particle distribution function $f$ using Dirac delta functions produced evolution equations similar to the standard LBM. The standard LBM introduces error terms in the hydrodynamic limit leading to issues such as lack of Galilean invariance in LBM simulations of fluids. Note that the dirac delta function has only two nonzero moments, the zero and first.  All higher moments are zero.  In this section, we explore using a different distribution function in the expansion of $f$ to overcome some of the challenges of the standard LBM.

\begin{figure}
    \centering
    \includegraphics[width=0.5\linewidth]{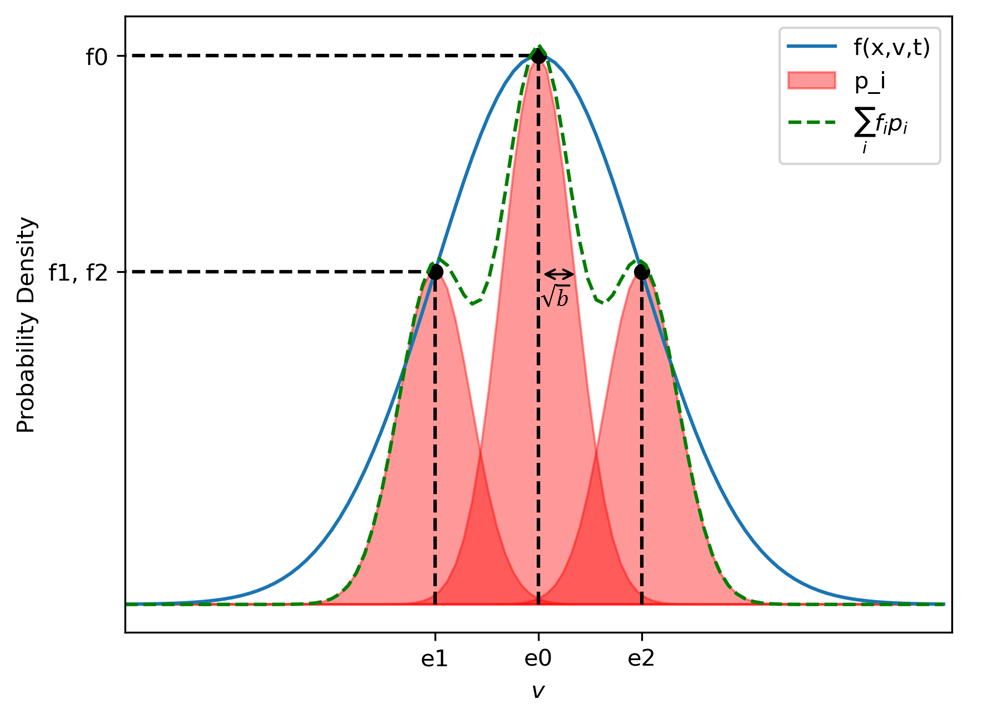}
    \caption{\footnotesize{A visual description of the new discretization method. The Maxwell-Boltzmann distribution is estimated by bump functions that not only have a mean but also have a variance.}}
    \label{fig:discretization_vis}
\end{figure}

Let's assume that
\begin{equation}\label{eq:new_discretization}
    f(\bm{x}, \bm{v}, t) = \sum_i f_i p(e_{i\alpha}, b_{\alpha \beta})
\end{equation}
where $p(e_{i\alpha}, b_{\alpha \beta})$ is a bump function (i.e. smooth and compactly supported) with mean ${\bf e}_i$ and variance $b_{\alpha \beta}$. The variance $b_{\alpha \beta}$ can have different properties but for now, we assume that it depends on space $\bm{x}$ and time $t$ but it does not depend on $i$. One can visualize the difference between the standard and new discretization as illustrated in Fig. \ref{fig:discretization_vis} for a one dimensional D1Q3 model. The blue solid line shows the particle distribution function $f$. Using the delta function, we sample 3 abscissae $e_{0}, e_{1}, e_{2}$ and the corresponding $f(e_i)$ become our discrete distributions $f_i$. The discrete velocities and distributions are indicated by black dashed lines. On the other hand, in the new discretization method, we estimate the continuous distribution function by the sum of bump functions similar to the ones shown in red in Fig. \ref{fig:discretization_vis}. The width of the red distributions is related to the $\sqrt{b}$. The estimated distribution function is realized by a green dashed line.  Note that there will be no need to specify an analytic form for $p$, just its first and second moments $(e_{i\alpha},b_{\alpha\beta})$.

\begin{figure}
    \centering
    \includegraphics[width=0.7\linewidth]{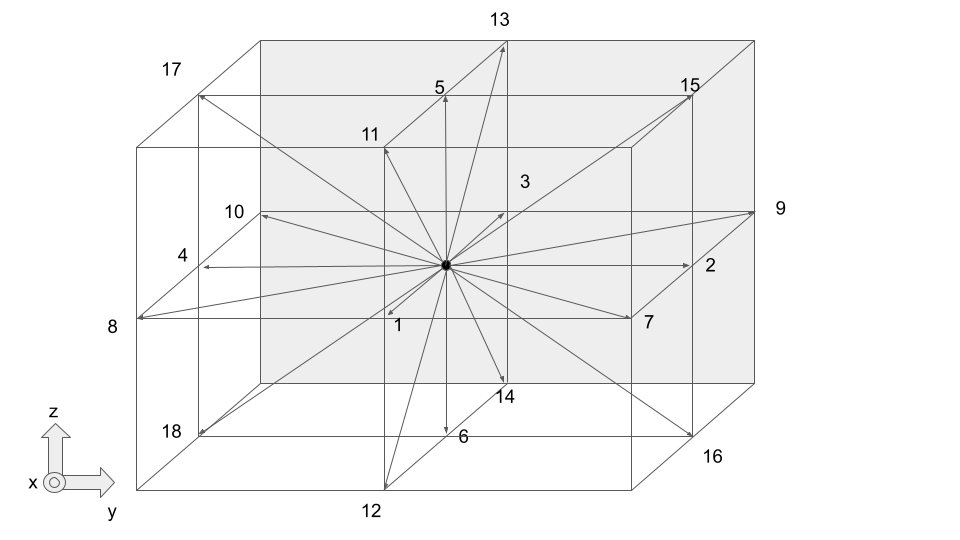}
    \caption{\footnotesize The D3Q19 lattice model is depicted. ${\bf e}_0=(0,0,0)$ is part of the velocity set but not labelled.}
    \label{fig:D3Q19}
\end{figure}

Now, based on this definition, in three dimensions we have an extra set of 6 variables (assuming $b_{\alpha \beta}$ is symmetric) which give us enough independent degrees of freedom at the 2nd and 3rd moments to actually set the equilibrium moments exactly to the moments of the MB distribution. If we plug Eq. (\ref{eq:new_discretization}) into the definition of the moments of the $f$, we get
\begin{align}
    &\int f d\bm{v} = \sum_i f_i \equiv \rho\\
    &\int f v_\alpha d\bm{v} = \sum_i f_i e_{i\alpha} \equiv \rho u_\alpha\\
    &\int f v_\alpha v'_\beta d\bm{v} d\bm{v}' = \sum_i f_i (e_{i\alpha} e_{i\beta} + b_{\alpha \beta}) \label{eq:new_2nd_moment}\\
    &\int f v_\alpha v'_\beta v''_\gamma d\bm{v} d\bm{v}' d\bm{v}'' = \sum_i f_i (e_{i\alpha} e_{i\beta} e_{i\gamma} + e_{i\alpha} b_{\beta \gamma} + e_{i\gamma} b_{\alpha \beta} + e_{i\beta} b_{\gamma \alpha})\label{eq:new_3rd_moment}.
\end{align}
Note that the moments of $f$ are not the discrete moments of $f_i$ (i.e. the sum over the product of $f_i$ and $e_i$'s) anymore. We want the moments of the equilibrium distribution function $f^{eq}$ to match the moments of the Maxwell-Boltzmann distribution, that is,
\begin{align}
    &\sum_i f_i^{eq} = \rho\label{eq:true_0th}\\
    &\sum_i f_i^{eq} e_{i\alpha} = \rho u_\alpha \\
    &\sum_i f_i^{eq} (e_{i\alpha} e_{i\beta} + b_{\alpha \beta}^{eq}) = P_{\alpha \beta} + \rho u_\alpha u_\beta\equiv \Pi_{\alpha\beta}^{eq}\\
    &\sum_i f_i^{eq} (e_{i\alpha} e_{i\beta} e_{i\gamma} + e_{i\alpha} b_{\beta \gamma}^{eq} + e_{i\gamma} b_{\alpha \beta}^{eq} + e_{i\beta} b_{\gamma \alpha}^{eq}) = \rho u_\alpha u_\beta u_\gamma+ P_{\beta \gamma} u_\alpha + P_{\alpha \beta} u_\gamma + P_{\gamma \alpha} u_\beta\nonumber\\
    & \qquad\qquad\qquad\qquad\qquad\qquad\qquad\qquad\qquad\quad \equiv Q_{\alpha\beta\gamma}^{eq} \label{eq:true_3rd}
\end{align}
and so on. Then the $f_i^{eq}$ are found by solving the set of linear equations above. Depending on how many moments we would like to match, we need different number of equations and a different lattice model. To match all the moments up to the third moment on a 3D lattice, we use the D3Q19 lattice shown in Fig. \ref{fig:D3Q19}. Generally, one would like to use fewer velocities for computational efficiency reasons. However, even with the new discretization method and in the presence of the $b$'s, it was not possible to find an isotropic solution using the D3Q15 model. Therefore, D3Q19 is the next best choice.

Solving the system of linear equations, we find solutions for $f_i^{eq}$ and $b^{eq}_{\alpha \beta}$ which are summarized in Appendix \ref{app:eq_dist}. The corresponding discrete moments of $f_i^{eq}$ are summarized below: 
\begin{align}
    &\sum_i f_i^{eq} = \rho \\
    &\sum_i f_i^{eq} e_{i\alpha} = \rho u_\alpha \\
    &\sum_i f_i^{eq} e_{i\alpha} e_{i\beta} = \frac{1}{3} \rho (2 u_\alpha u_\beta + c^2 \delta_{\alpha \beta})\\
    &\sum_i f_i^{eq} e_{i\alpha} e_{i\beta} e_{i\gamma} = \frac{1}{3}\rho c^2 (\delta_{\beta \gamma} u_\alpha + \delta_{\alpha \beta} u_\gamma + \delta_{\gamma \alpha} u_\beta) \label{eq:3rd_moment_fi}
\end{align}
And, the $b_{\alpha \beta}^{eq}$ are
\begin{equation} \label{eq:b_to_P}
    b_{\alpha \beta}^{eq} = P_{\alpha \beta}/\rho + \frac{1}{3}(u_\alpha u_\beta - c^2).
\end{equation}
Not surprisingly, we see that the discrete moments of $f_i^{eq}$ are different from those of the standard model as some of the distribution is carried by the $b_{\alpha \beta}$ terms.  Also, while the third {\it discrete} moment of the $f_i$ in Eq.(\ref{eq:3rd_moment_fi}) is the same as that of the standard model, the full third moment of the entire distribution which includes the contribution from the $b$'s in Eq.(\ref{eq:true_3rd}) is, in fact, what is necessary to match the third moment of the Maxwell-Boltzmann distribution. 

The next step is to find the evolution equations for the new model. Similar to what we did in section \ref{subsec:delta}, we start with the {\it continuous} Boltzmann equation as our foundation and integrate over the velocity space to get the evolution equation for the lowest moment to be
\begin{align}
    \int_{\Omega_v} \left(\partial_t f + v_\alpha \partial_\alpha f\right) d\bm{v} &=  \sum_i \left(\partial_t \int_{\Omega_v}  f_i p_i d\bm{v} + \int_{\Omega_v} v_\alpha \partial_\alpha  f_i p_i d\bm{v} \right)\nonumber \\
    &=  \sum_i \left(\partial_t  f_i + e_{i\alpha} \partial_\alpha  f_i \right)
\end{align}
where $p_i = p(e_{i\alpha}, b_{\alpha \beta})$. Therefore,
\begin{equation}
    \sum_i \left\{\partial_t  f_i + e_{i\alpha} \partial_\alpha  f_i = -\frac{1}{\tau}(f_i - f_i^{eq})\right\}\label{eq:new_algo_zeroth}.
\end{equation}
We do not see a difference between the standard LBM and the new model at the zeroth moment as the $b_{\alpha \beta}$ terms only appear in the second moments and higher. However, for the first and second moments, multiplying the continuous Linearized Boltzmann equation in Eq. (\ref{eq:LinBoltz}) by powers of $v_\beta$ ({\it not} $e_{i\beta}$)  and then integrating over $v$, the evolution equations are found to be
\begin{equation}\label{eq:new_evol_1st}
    \sum_i \left\{\partial_t  (f_i e_{i\beta}) + \partial_\alpha  (f_i (e_{i\alpha} e_{i\beta}+b_{\alpha \beta}))= -\frac{1}{\tau}(f_i e_{i\beta}- f_i^{eq}e_{i\beta})\right\},
\end{equation}
\begin{align}\label{eq:new_evol_2nd}
    \sum_i \left\{\partial_t  (f_i (e_{i\beta}e_{i\gamma}+b_{\beta \gamma})) + \partial_\alpha  (f_i (e_{i\alpha} e_{i\beta} e_{i\gamma} \right. &+
    e_{i\alpha} b_{\beta \gamma} + e_{i\gamma} b_{\alpha \beta} + e_{i\beta} b_{\gamma \alpha}))\nonumber \\ 
    & = \left. -\frac{1}{\tau}\left(f_i (e_{i\beta}e_{i\gamma}+b_{\beta \gamma})- f_i^{eq} (e_{i\beta}e_{i\gamma}+b^{eq}_{\beta \gamma})\right)\right\} .
\end{align}
As can be seen, the evolution equation for the higher moments are {\it different} from the standard LBM. However, it is possible to rearrange the above equations to be able to evolve the new evolution equations using a standard LB algorithm (with collisions in moment space) coupled with additional equations to evolve the $b_{\alpha\beta}$ This can be done by separating the evolution equations for the first and second moments of $f_i$ and $b_{\alpha \beta}$:
\begin{align}
	&\sum_i \left\{\partial_t (f_ie_{i\beta}) + \partial_{\alpha} (f_i e_{i\alpha}e_{i\beta}) = -\frac{1}{\tau}(f_i -f_i^{eq})-\partial_{\alpha} (f_i b_{\alpha \beta})\right\},\label{eq:new_algo_1st}\\
	&\sum_i\left\{\partial_t \left(f_i \left(e_{i\beta}e_{i\gamma} - \frac{c^2}{3}\delta_{\alpha \beta}\right)\right) + \partial_{\alpha} \left(f_i e_{i\alpha} \left(e_{i\beta}e_{i\gamma} - \frac{c^2}{3}\delta_{\alpha \beta}\right)\right)\right. \nonumber\\
	&\qquad \qquad \qquad \left.= -\frac{1}{\tau}(f_i -f_i^{eq})\left(e_{i\beta}e_{i\gamma} - \frac{c^2}{3}\delta_{\alpha \beta}\right) - \partial_{\alpha} (f_i (e_{i\gamma} b_{\alpha \beta}+e_{i\beta} b_{\gamma \alpha}))\right\},\label{eq:new_algo_2nd}\\	
	&\sum_i\left\{\partial_t \left(f_i \left(b_{\beta \gamma}+\frac{c^2}{3}\delta_{\beta\gamma}\right)\right) + \partial_{\alpha} \left(f_i e_{i\alpha} \left(b_{\beta \gamma}+\frac{c^2}{3}\delta_{\beta\gamma}\right)\right)= -\frac{1}{\tau}(f_i b_{\beta \gamma} - f_i b^{eq}_{\beta \gamma}) \right\}\label{eq:b_evo}
\end{align}
Here we have split Eq.(\ref{eq:new_evol_2nd}) into equations for the evolution of the second discrete moments of $f_i$ and an equation for the evolution of the $b$'s (if both of these equations are true then their sum in Eq.(\ref{eq:new_evol_2nd}) will also be true). The extra terms in Eq. (\ref{eq:new_algo_1st}) and (\ref{eq:new_algo_2nd}) that appear in form of derivatives of $b$ couple the evolution of $f_i$ and $b$'s. These can be treated as part of the forcing terms in a standard Lattice Boltzmann algorithm. When interpreted as a forcing term in a standand Lattice Boltzmann algorithm, and taking into account the relation between $b$ and the pressure Eq.(\ref{eq:b_to_P}), the last term in Eq.(\ref{eq:new_algo_1st}) is similar to the non-ideal pressure forcing terms suggested by \cite{He1998b,He2002}. However, the inclusion of the velocity term in Eq.(\ref{eq:b_to_P}), the corresponding change in the second discrete moment of the $f_i$, and the corresponding second moment forcing in Eq.(\ref{eq:new_algo_2nd}) make this unique.  In the next section, we discuss the derivation of the discrete space and time LB equation from the discrete Boltzmann equation. 
\subsection{Space and Time Discretization}\label{subsec:time_discretization}
Let us start with the discrete-velocity linearized Boltzmann equation in the most general form
\begin{equation}
	\partial_t f_i + e_{i\alpha} \partial_{\alpha} f_i = -\mathcal{L}_{ij}(f_j -f_j^{eq}) + \Phi_i
\end{equation}
where $f_i$ is the discrete-velocity distribution function, and $\bm{e}_{i}$ are the discrete velocities that $f_i$ travels to neighbouring sites by. $\mathcal{L}_{ij}$ is the collision operator, and $\Phi_i$ is the forcing term. Our derivation will be similar to that in Reference \cite{Nash2008} for the standard LBM.  For sake of brevity, we call the RHS of the above equation $C_i$
\begin{equation}
	-\mathcal{L}_{ij}(f_j -f_j^{eq}) + \Phi_i  := C_i (\bm{x}, t)
\end{equation}
One can solve the discrete-velocity Boltzmann equation using the method of characteristics
\begin{equation}
	\frac{df_i}{dt} = C_i
\end{equation}
where $d/dt$ is the complete derivative with respect to $t$. Taking the integral of both sides, we get
\begin{equation}
	f_i(\bm{x}+\bm{e}_i\Delta t, t+\Delta t) - f_i(\bm{x}, t) = \int_0^{\Delta t} C_i(\bm{x}+\bm{e}_i s, t+s) ds
\end{equation}
Now, if we perform the trapezoidal rule to solve the integral, we find
\begin{equation}\label{eq:trapLB}
	f_i(\bm{x}+\bm{e}_{i}\Delta t, t+\Delta t) - f_i(\bm{x}, t) = \frac{C_i(\bm{x}, t)+C_i(\bm{x}+\bm{e}_{i}\Delta t, t+ \Delta t)}{2}\Delta t + \mathcal{O}(\Delta t^3)
\end{equation}
Reorganizing (\ref{eq:trapLB}), we can write
\begin{equation}\label{eq:fbarLB}
	\bar{f}_i(\bm{x}+\bm{e}_{i}\Delta t, t+\Delta t) = \bar{f}_i(\bm{x}, t) + C_i(\bm{x}, t)\Delta t
\end{equation}
where
\begin{equation}\label{eq:fbardef}
	\bar{f}_i(\bm{x}, t) = f_i(\bm{x}, t)-\frac{C_i(\bm{x}, t)}{2}\Delta t.
\end{equation}
Eq. (\ref{eq:fbarLB}) is the LB equation in terms of the auxiliary distributions $\bar{f}_i$. The fact that the LB can be recovered using the method above shows that the (fully) discretized LB equation is actually accurate up to the second order \cite{Kruger2016book,Dellar2001}. To benefit from this higher order accuracy, one can use Eq. (\ref{eq:fbarLB}) as the LB evolution equation with the caveat that the macroscopic moments need to be corrected to match physical parameters. This is due to the fact that it is the moments of $f_i$ and not $\bar{f}_i$ that are linked with the macroscopic parameters (moments). We get to these correction terms soon but first, we have to make sure that Eq. (\ref{eq:fbarLB}) is all in terms of $\bar{f}_i$ for it to be used as an evolution equation. This means that $C_i$ has to be transformed. Plugging Eq. (\ref{eq:fbardef}) into the definition of $C_i$, we get
\begin{align}
		 C_i &= -\mathcal{L}_{ij}(f_j -f_j^{eq}) + \Phi_i \nonumber\\
		 	 & = -\frac{\Delta t}{2}\mathcal{L}_{ij} C_j-\mathcal{L}_{ij}(\bar{f}_j -f_j^{eq}) + \Phi_i
\end{align}
Taking all the terms with $C_i$ to one side and performing some linear algebra, we get
\begin{equation}
	\left(\delta_{ij} +\frac{\Delta t}{2}\mathcal{L}_{ij} \right)C_j =-\mathcal{L}_{ij}(\bar{f}_j -f_j^{eq}) + \Phi_i,
\end{equation}
and finally, we have $C$ in terms of $\bar{f}_i$:
\begin{equation}
	C_j =\left(\mathbb{I} +\frac{\Delta t}{2}\mathcal{L} \right)_{ji}^{-1}(-\mathcal{L}_{ik}(\bar{f}_k -f_k^{eq}) + \Phi_i).
\end{equation}
Therefore, Eq. (\ref{eq:fbarLB}) can be rewritten fully in terms of $\bar{f}_i$ as
\begin{equation}\label{eq:fbarLB_full}
	\bar{f}_i(\bm{x}+\bm{e}_{i}\Delta t, t+\Delta t) = \bar{f}_i(\bm{x}, t) + \left[\left(\mathbb{I} +\frac{\Delta t}{2}\mathcal{L} \right)_{ij}^{-1}\left(-\mathcal{L}_{jk}(\bar{f}_k -f_k^{eq}) + \Phi_j\right)\right] \Delta t.
\end{equation} 
For the single-relaxation time model, this would take the form
\begin{equation}\label{eq:bbardef}
    \bar{f}_i(\bm{x}+\bm{e}_{i}\Delta t, t+\Delta t) = \bar{f}_i(\bm{x}, t) + \frac{\Delta t}{\tau+\frac{\Delta t}{2}}\left(-(\bar{f}_k -f_k^{eq}) + \tau \Phi_j\right).
\end{equation}

The hydrodynamic variables are found from the moments of $f_i$. However, the evolution equation above is in terms of $\bar{f}_i$. Therefore, the macroscopic equations obtained from the evolution equation will have moments of $\bar{f}_i$. We can use Eq. (\ref{eq:fbardef}) to correct the moments:
\begin{align}\label{eq:correction_terms}
    \rho &= \sum_i \bar{f}_i \\
    \rho u_\beta &= \sum_i \bar{f}_i e_{i\beta} + \sum_i \Phi_i e_{i\beta}\frac{\Delta t}{2} \nonumber\\
    &= \sum_i \bar{f}_i e_{i\alpha} -  \partial_\alpha (\rho b_{\alpha \beta})\frac{\Delta t}{2}
\end{align}
where
\begin{align}
    \sum_i \Phi_i &= 0,\\
    \sum_i \Phi_i e_{i\beta} &= -\partial_\alpha (\rho b_{\alpha \beta}),\\
    \sum_i \Phi_i e_{i\beta}e_{i\gamma} &=  - \sum_i \partial_{\alpha} (f_i (e_{i\gamma} b_{\alpha \beta}+e_{i\beta} b_{\gamma \alpha})),\\
    & = - \partial_{\alpha}(\rho u_\gamma b_{\alpha \beta} + \rho u_\beta b_{\gamma \alpha} ).\label{eq:2nd_mom_corr}
\end{align}
If the velocity was not under the derivative, the second moment forcing term would be similar to that seem in typical forcing schemes \cite{Guo2002b}.  However, having the velocity inside the derivative will be vital for the model to eliminate the terms that destroy Galilean invariance.
Since the forcing term $\Phi_i$ involves $b_{\alpha \beta}$ terms which are subject to collision and time evolution, we have to account for this variation with time.
Similar to $f_i$, we define an auxiliary $\bar{b}_{\alpha \beta}$ which can be obtained starting from Eq. \ref{eq:b_evo} and following the same steps we used to derive $\bar{f}_i$:
\begin{equation}
    \bar{b}_{\alpha \beta} (\bm{x}, t)= b_{\alpha \beta} (\bm{x}, t)- \frac{\Delta t}{2} B_{\alpha \beta}(\bm{x}, t)
\end{equation}
where
\begin{equation}
    B_{\alpha \beta}(\bm{x}, t) = -\frac{1}{\tau}(b_{\alpha \beta} - b^{eq}_{\alpha \beta}).
\end{equation}
Following the same procedure as above, the fully discrete evolution of the $\bar{b}_{\alpha \beta}$ is found to be
\begin{equation}
     \bar{f}_i \bar{b}_{\beta \gamma} (\bm{x}+\bm{e}_{i}\Delta t, t+\Delta t) = \bar{f}_i \bar{b}_{\beta \gamma}(\bm{x}, t) + \Delta t \bar{f}_i B_{\beta \gamma}(\bm{x}, t).
\end{equation}
As with the $C_i$, in application we need the $B_{\beta \gamma}$ terms as a function of $\bar{b}_{\beta \gamma}$ which can be found in analgous manner to obtaining the $C_i$ in terms of $\bar{f}_i$ to get
\begin{equation}
	B_{\beta \gamma}(\bm{x}, t)=-\frac{1}{\tau+\frac{\Delta t}{2}}\left(\bar{b}_{\beta \gamma}(\bm{x}, t)-b^{eq}_{\beta \gamma}(\bm{x}, t)\right).
\end{equation}
In order to obtain $\bar{b}_{\beta \gamma}$ at $t'= t+\Delta t$ using this expression, it is easier to shift positions so that the left hand side is evaluated at $\bm{x}$ and complete the sum over $i$ to get:
\begin{equation}
	\bar{b}_{\beta \gamma}(\bm{x}, t')=\frac{1}{\rho(\bm{x}, t')}\sum_i\left\{ \bar{f}_i \left( \bar{b}_{\beta \gamma} + \Delta t B_{\beta \gamma} \right) \right\}_{(\bm{x}-\bm{e}_{i}\Delta t, t'-\Delta t)},
\end{equation}
where all variables in term $i$ in the sum are evaluated at $(\bm{x}-\bm{e}_{i}\Delta t, t'-\Delta t)$.

The $\mathcal{O}(\Delta t)$ correction to $\bm{u}$ in Eq.(\ref{eq:correction_terms}) can then be written in terms of the auxiliary parameters:
\begin{align}
    \partial_\alpha (\rho b_{\alpha \beta}) &= \partial_\alpha (\rho \bar{b}_{\alpha \beta}+ \frac{\Delta t}{2}B) \\
    &= \partial_\alpha \left[\rho \bar{b}_{\alpha \beta}- \frac{\Delta t/2}{\tau + \Delta t/2} \left(\rho \bar{b}_{\alpha \beta}-\rho b^{eq}_{\alpha \beta}\right)\right].
\end{align}
In principle this is an implicit equation as $b^{eq}_{\alpha \beta}$ contains a quadratic velocity term $u_\alpha u_\beta$ in Eq.(\ref{eq:b_to_P}).  However, if one uses the uncorrected $\bm{u}$ as an approximation in $b^{eq}_{\alpha \beta}$ here, the resulting ``corrected" $\bm{u}$ will only differ from the exact solution to the implicit equation by a term of $\mathcal{O}(\Delta t^3)$, which is comparable to other discretization errors made in the trapezoidal rule.  The second moment correction in Eq.(\ref{eq:2nd_mom_corr}) can be evaluated similarly. The spatial derivatives in the moment ``corrections" must be evaluated using a finite-difference scheme.  For consistency with the rest of the lattice Boltzmann algorithm we chose the isotropic finite difference schemes of reference \cite{THAMPI20131}.  For the applications shown in Section \ref{sec:application} all simulations are done on a D3Q19 lattice.
\subsection{Chapman-Enskog Analysis}\label{subsec:Chapman}
Here we perform the Chapman-Enskog analysis to show that our model recovers the governing macroscopic equations, i.e. Navier-Stokes-Fourier equations. We do this in a slightly non-traditional way. Typically the Chapman-Enskog expansion is applied to the LB equation. We showed in section \ref{subsec:time_discretization} that the discrete space and time equations are consistent with the continuous (in space and time but still discrete velocity) Boltzmann Equation to at least second order in time. Thus, we can perform our Chapman-Enskog on the discrete velocity Boltzmann equation (continuous in time and space) and its moments, Eqs. (\ref{eq:new_algo_zeroth}), (\ref{eq:new_evol_1st}), and (\ref{eq:new_evol_2nd}) instead.

The Chapman-Enskog analysis is an expansion about equilibrium.  We assume a small deviation from the equilibrium such that $f_i \approx f_i^{eq} + \epsilon f^{(1)}$ where $\epsilon$ is small.  At the lowest order, we substitute $f_i \approx f_i^{eq}$ into Eqs. (\ref{eq:new_algo_zeroth}), (\ref{eq:new_evol_1st}), and (\ref{eq:new_evol_2nd}) and complete the sum over $i$ using the equlibrium moments given in Eqs.(\ref{eq:true_0th})-(\ref{eq:true_3rd}), we get the lowest order evolution equations:
\begin{align}
	\partial_t \rho + \partial_{\alpha}(\rho u_{\alpha}) &= 0 + \mathcal{O}(\epsilon) \label{eq:continuity_zeroth}\\
	\partial_t (\rho u_{\beta}) + \partial_{\alpha}\Pi_{\alpha \beta}^{eq} &= 0 + \mathcal{O}(\epsilon)\label{eq:momentum_zeroth}\\
	\partial_t \Pi_{\beta \gamma}^{eq}  + \partial_{\alpha}Q_{\alpha \beta \gamma}^{eq} &= 0+ \mathcal{O}(\epsilon).
\end{align}
To get the next order in the expansion, we use $f_i \approx f_i^{eq} + \epsilon f^{(1)}$ and keep all terms up to $\mathcal{O}(\epsilon)$ to get:
\begin{align}
	\partial_t \rho + \partial_{\alpha}(\rho u_{\alpha}) &= 0 + \mathcal{O}(\epsilon^2)\label{eq:zeroth_ChE}\\
	\partial_t (\rho u_{\beta}) + \partial_{\alpha}\Pi_{\alpha \beta}^{eq} &= -\partial_{\alpha} \Pi_{\alpha \beta}^{(1)}+ \mathcal{O}(\epsilon^2) \label{eq:first_ChE}\\
	\partial_t \Pi_{\beta \gamma}^{eq}  + \partial_{\alpha}Q_{\alpha \beta \gamma}^{eq} &= -\frac{1}{\tau} \Pi_{\beta \gamma}^{(1)} + \mathcal{O}(\epsilon^2)\label{eq:second_ChE}
\end{align}
where $\Pi_{\alpha \beta}^{(1)}$ is the deviation of the second moment from $\Pi_{\alpha \beta}^{eq}$ due to $f^{(1)}$. From above, we see that one can find the non-equilibrium moments at a certain level by solving the equation one order higher using the equilibrium moments. For the momentum equation, we have to find $\Pi_{\alpha \beta}^{(1)}$ using equation (\ref{eq:second_ChE}) involving the equilbrium third-moment.  We start plugging the equilibrium moments into the left-hand side:
\begin{align}
	\partial_t \Pi_{\beta \gamma}^{eq}  + \partial_{\alpha} Q_{\alpha \beta \gamma}^{eq} 
	&= \partial_t (P_{\beta \gamma} + \rho u_{\beta} u_{\gamma}) + \partial_{\alpha}(P_{\alpha \beta} u_{\gamma}+P_{\gamma \alpha} u_{\beta}+P_{\beta \gamma} u_{\alpha}+ \rho u_{\alpha} u_{\beta} u_{\gamma}) \\
	&= \partial_t (P_{\beta \gamma}) + \partial_t (\rho u_{\beta} u_{\gamma}) + \partial_{\alpha}(P_{\alpha \beta} u_{\gamma}+P_{\gamma \alpha} u_{\beta}+P_{\beta \gamma} u_{\alpha}) + \partial_{\alpha}(\rho u_{\alpha} u_{\beta} u_{\gamma}).
\end{align}
Using Eq. (\ref{eq:continuity_zeroth}) and Eq. (\ref{eq:momentum_zeroth}) to drop higher order terms, and the product rule
\begin{equation}
    \partial (abc) = a \partial (bc) + b \partial (ac) - ab \partial (c),
\end{equation}
we can write
\begin{align}
	\partial_t \Pi_{\beta \gamma}^{eq}  + \partial_{\alpha} Q_{\alpha \beta \gamma}^{eq} 
	&= \partial_t (P_{\beta \gamma}) + \partial_t (\rho u_{\beta}) u_{\gamma} + \partial_t (\rho u_{\gamma}) u_{\beta} - (\partial_t \rho) u_{\beta} u_{\gamma}\nonumber\\
	&\qquad \qquad \qquad + \partial_{\alpha}(P_{\alpha \beta} u_{\gamma}+P_{\gamma \alpha} u_{\beta}+P_{\beta \gamma} u_{\alpha}) + \partial_{\alpha}(\rho u_{\alpha} u_{\beta} u_{\gamma}) \nonumber\\
	&= \partial_t (P_{\beta \gamma}) - \partial_{\alpha} (P_{\alpha \beta}) u_{\gamma} - \partial_{\alpha} (P_{\alpha \gamma}) u_{\beta} + \partial_{\alpha}(P_{\alpha \beta} u_{\gamma}+P_{\gamma \alpha} u_{\beta}+P_{\beta \gamma} u_{\alpha})
\end{align}
Assuming that $P_{\alpha \beta} = P(\rho)$, we have
\begin{align}
	\partial_t \Pi_{\beta \gamma}^{eq}  + \partial_{\alpha} Q_{\alpha \beta \gamma}^{eq}
		&= \partial_{\rho} (P_{\beta \gamma}) \partial_t \rho - \partial_{\alpha} (P_{\alpha \beta}) u_{\gamma} - \partial_{\alpha} (P_{\alpha \gamma}) u_{\beta}  + \partial_{\alpha}(P_{\alpha \beta} u_{\gamma} +P_{\gamma \alpha} u_{\beta}+P_{\beta \gamma} u_{\alpha}) \\
		&= (P_{\beta \gamma}-\rho \partial_{\rho} (P_{\beta \gamma}))  \partial_{\alpha}  u_{\alpha} + P_{\alpha \beta}\partial_{\alpha} u_{\gamma} +P_{\alpha \gamma} \partial_{\alpha} u_{\beta}. 
	\end{align}
Substituting $P_{\alpha \beta} = \rho T \delta_{\alpha \beta}$, gives
\begin{align}
	\partial_t \Pi_{\beta \gamma}^{eq}  + \partial_{\alpha} Q_{\alpha \beta \gamma}^{eq}&= (\rho T \delta_{\beta \gamma}-\rho \partial_{\rho} (\rho T \delta_{\beta \gamma}))  \partial_{\alpha}  u_{\alpha} \nonumber\\
	& \qquad \qquad  + \rho T \delta_{\alpha \beta}\partial_{\alpha} u_{\gamma} +\rho T \delta_{\alpha \gamma} \partial_{\alpha} u_{\beta}\\
	&= \rho T \partial_{\beta} u_{\gamma} +\rho T \partial_{\gamma} u_{\beta}. \label{eq:ChE_momentum}
\end{align}
Note that the temperature $T$ here is in energy units and has the following relation with the physical temperature
\begin{equation}
T = \frac{R\theta}{M}
\end{equation}
where $\theta$ is the temperature, $R$ is the universal gas constant, and $M$ is the molar mass of the specific matter.

From Eq. (\ref{eq:second_ChE}), Eq. (\ref{eq:ChE_momentum}) is equal to -$\frac{1}{\tau} \Pi_{\beta \gamma}^{(1)}$ which can then be substituted into Eq. (\ref{eq:first_ChE}) to get the final momentum equation
\begin{equation}\label{eq:LB_NSE}
	\partial_t (\rho u_{\beta}) + \partial_{\alpha}\Pi_{\alpha \beta}^{eq} = \partial_{\alpha} \left(\eta \bigg(\partial_{\beta} u_{\gamma} + \partial_{\gamma} u_{\beta} \bigg)\right)
\end{equation}
where the viscosity $\eta = \rho T \tau$. If we rearrange the above equation, we get
\begin{equation}
	\partial_t (\rho u_{\beta}) + \partial_{\alpha}\Pi_{\alpha \beta}^{eq} = \partial_{\alpha} \left(\eta \bigg(\partial_{\beta} u_{\gamma} + \partial_{\gamma} u_{\beta} -\frac{2}{3}\partial_{\delta} u_\delta \delta_{\beta \gamma} \bigg) + \frac{2}{3} \eta \partial_{\delta} u_\delta \delta_{\beta \gamma}\right)
\end{equation}
which is the exact Navier-Stokes equation with a bulk viscosity of $\frac{2}{3} \eta$.

\section{Application}\label{sec:application}
The governing macroscopic equations of our model were derived in section \ref{subsec:Chapman}. In this section, we measure the physical properties of our fluid. We also compare these results to theoretical predictions to validate our model.
\subsection{Poiseuille Flow}
In this section, we measure the shear viscosity in a Poiseuille flow and compare it to the predicted viscosity $\eta = \rho T \tau$. A Poiseuille flow can be generated by applying a body force or a pressure gradient along the channel between two parallel stationary walls. In our case, we apply a body force acceleration of $a = 10$ m/s$^2$ $= 0.00001 \frac{\mu m}{\mu s^2}$ in the y direction. A fluid with viscosity $\eta = 0.8$ cP $=0.8 \frac{\text{pg}}{\mu \text{m} \mu \text{s}}$ and density $\rho = 1.0 \frac{\text{g}}{\text{cm}^3} = 1.0 \frac{\text{pg}}{\mu \text{m}^3}$ is used. The LB timestep and lattice spacing are $\Delta t = 2 \mu \text{s}$ and $\Delta x = 2 \mu \text{m}$, respectively. The theoretical velocity profile when a body force acceleration $a$ in the y direction is applied can be found from the momentum equation (Eq. \ref{eq:LB_NSE}):
\begin{equation}\label{eq:poiseuille_momentum}
	\partial_t (\rho u_{\beta}) + \partial_{\alpha}(\rho u_{\alpha} u_{\beta}) = - \partial_{\beta}(\rho T) + \eta \partial_{\alpha} \bigg(\partial_{\beta} u_{\alpha} + \partial_{\alpha} u_{\beta} \bigg) + \rho a_\beta.
\end{equation}

In steady-state, nothing changes with time. Thus, $\partial_t \ = 0$. The walls of the channel are set up with normals in the z direction and a half-step bounce back boundary condition is applied at the walls \cite{Ladd1994}. Periodic boundary conditions are used in the $x$ and $y$ directions. Therefore, everything is translationally invariant in these directions, i.e. $\partial_x \ = \partial_y \  = 0$. There is also no driving force applied in the x direction resulting in no velocity in that direction, i.e. $u_x = 0$. The walls and conservation of mass impose $u_z = 0$. Considering all the above conditions and an incompressible regime ($Ma = u_{max}/c_s < 0.35$), Eq. \ref{eq:poiseuille_momentum} transforms into
\begin{equation}
	0 = 0 + \eta \partial_{z} \bigg(\partial_{y} u_{z} + \partial_{z} u_{y} \bigg) + \rho a_y.
\end{equation}
which in turn has a solution of form
\begin{equation}
	u_y(z) = -\frac{1}{2\eta}\rho a_y z(z-H)
\end{equation}
where $a_y$ is the y component of the acceleration (rest of them are zero), and $H$ is the height of the channel in the z-direction. 

\begin{figure}[!htpb]
	\centering
	\begin{subfigure}[t]{0.3285\textwidth}
		\includegraphics[width=\textwidth]{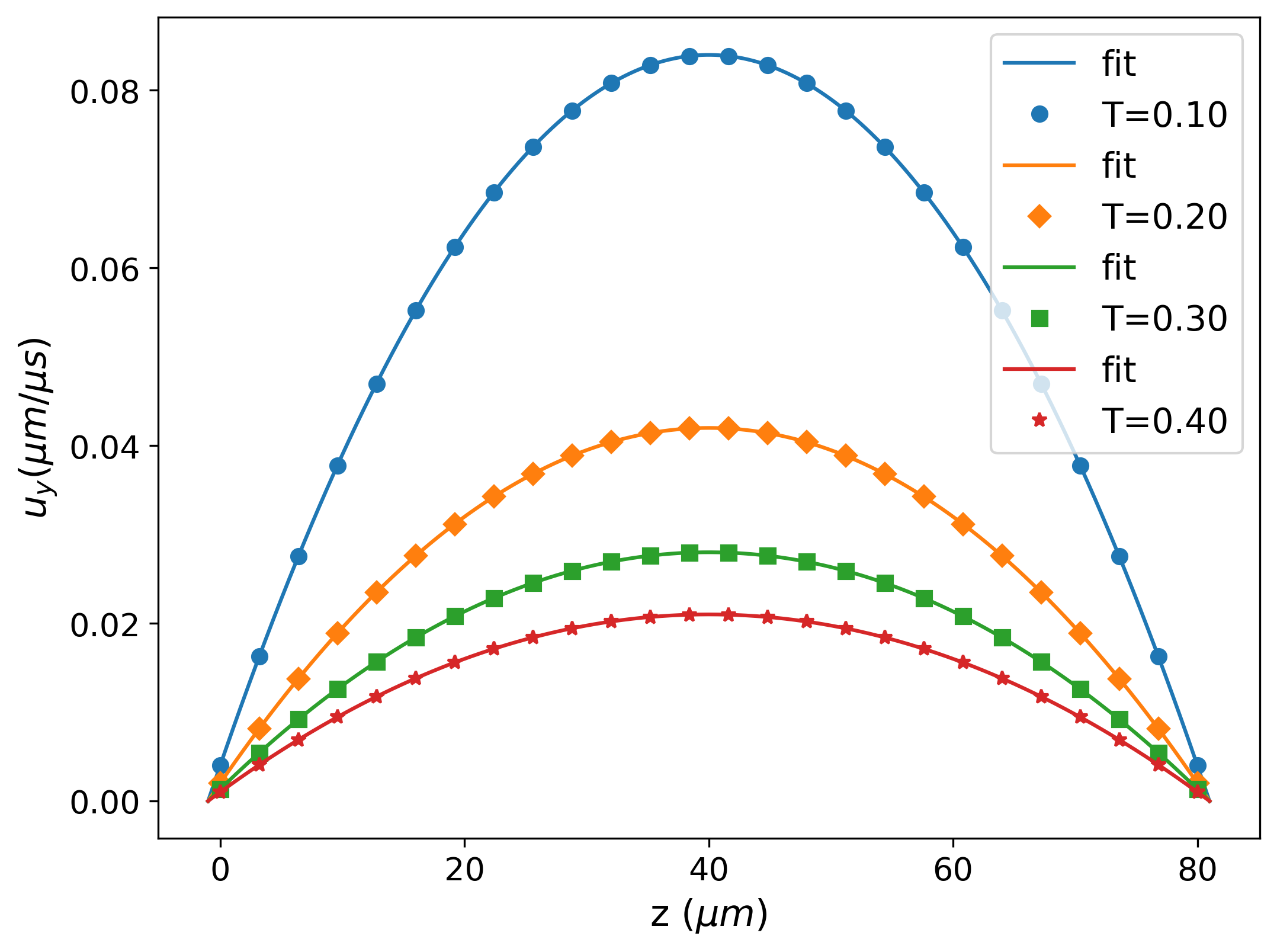}
		\caption{}
	\end{subfigure}
    	\begin{subfigure}[t]{0.3285\textwidth}
		\includegraphics[width=\textwidth]{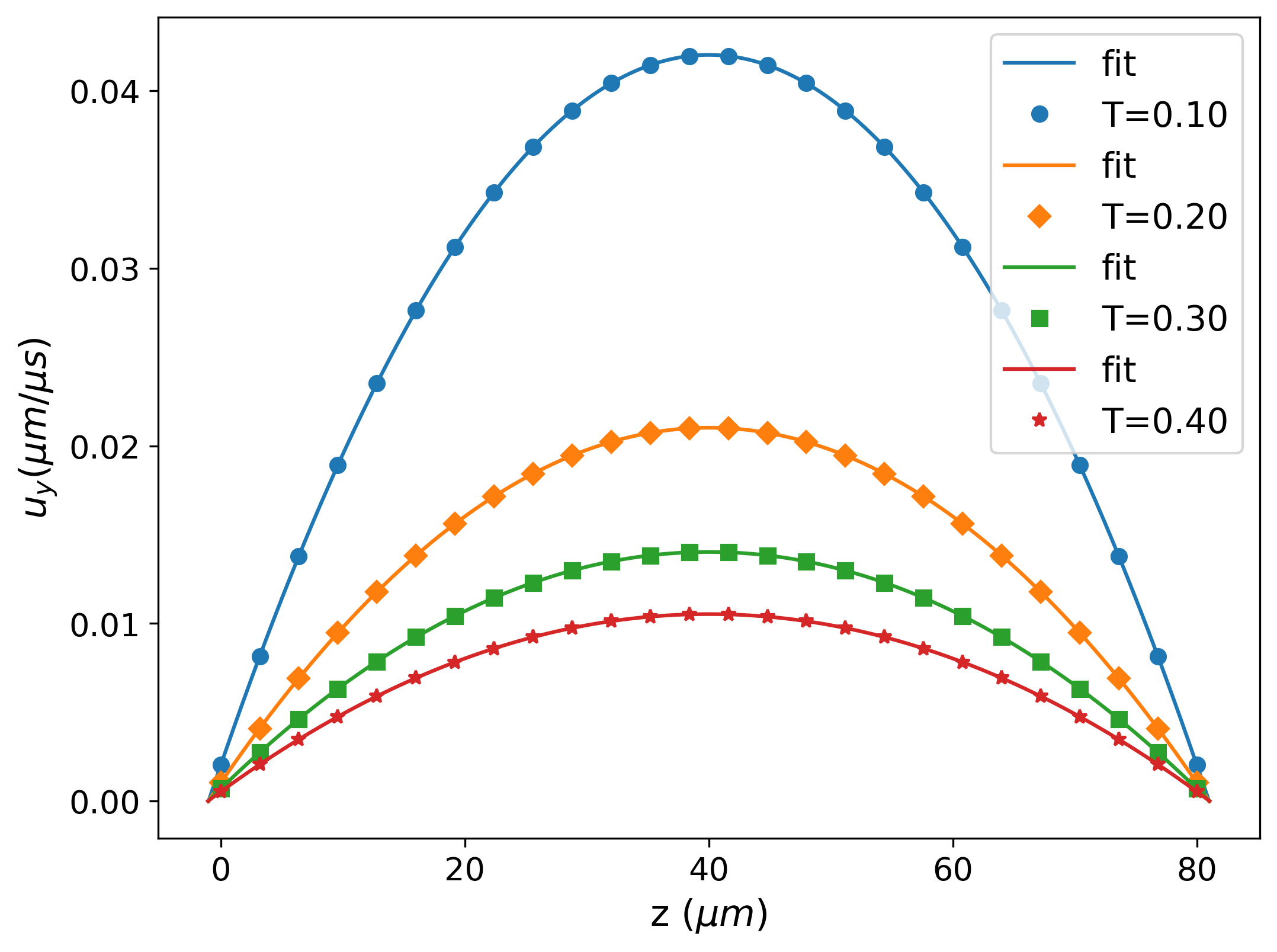}
		\caption{}
	\end{subfigure}
	\begin{subfigure}[t]{0.3285\textwidth}
		\includegraphics[width=\textwidth]{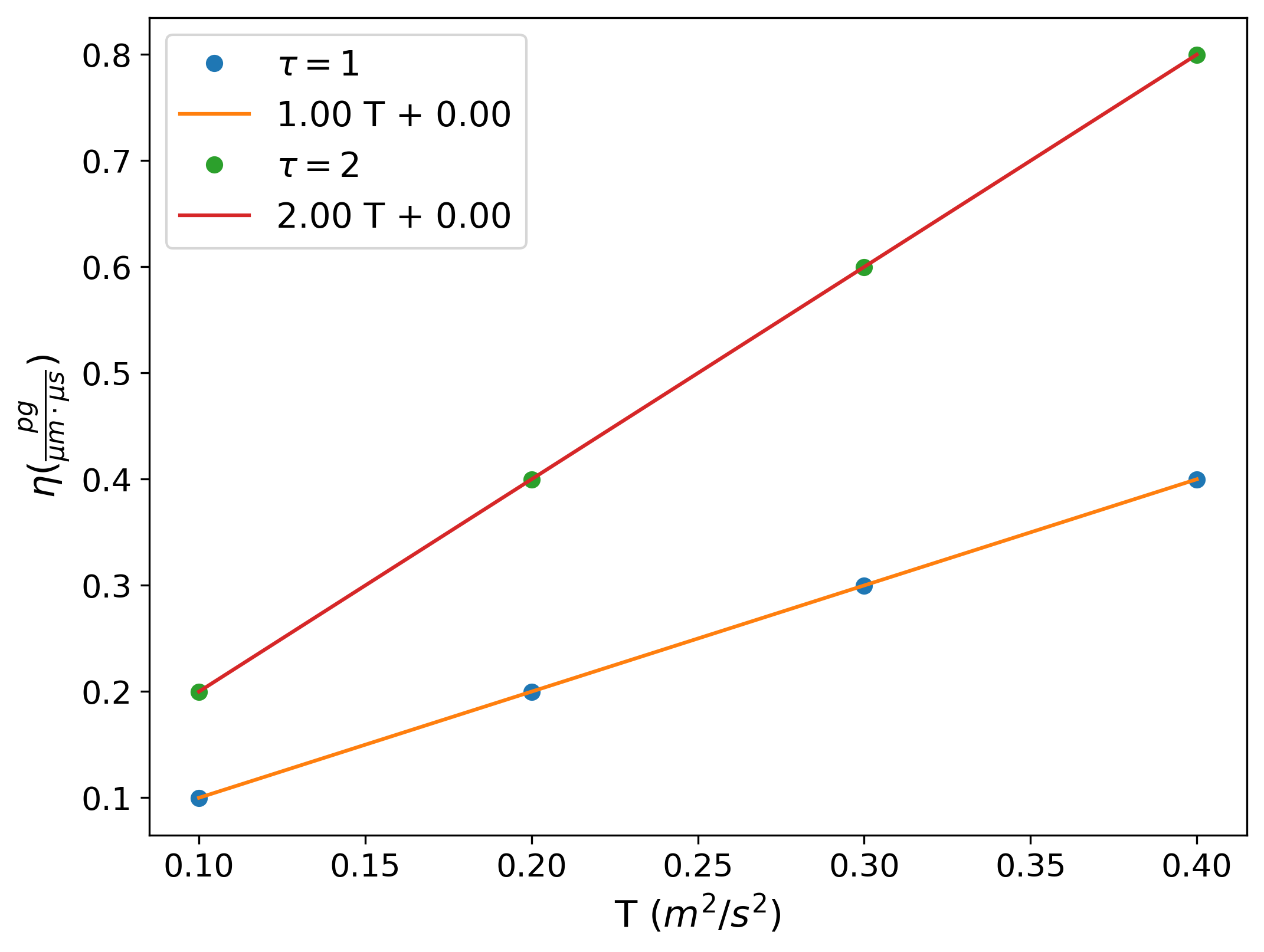}
		\caption{}
	\end{subfigure}
	\caption{\footnotesize{The velocity profile for a Poiseuille flow at different temperatures are shown for $\tau = 1 \mu \text{s}$ (a) and $\tau = 2 \mu \text{s}$ (b). (c) shows the measured viscosity as a function of temperature. A linear relation between viscosity and temperature is obtained. The slope of the line is $\rho \tau$ which agrees with the theoretical prediction.}}
	\label{fig:Poiseuille}
\end{figure}

Fig. \ref{fig:Poiseuille}(a) and (b) illustrate the $u_y$ profile for fluids at different temperatures and two relaxation times $\tau = 1 \mu \text{s}$ and $\tau = 2 \mu \text{s}$, respectively. For these simulations, we observe that the fluid reaches larger velocities at higher temperatures. Based the Chapman-Enskog expansion, we expect the viscosity and temperature to be related by
\begin{equation}\label{eq:visc-T_relation}
	\eta = \rho T \tau
\end{equation}
where $\tau$ is the relaxation time. To confirm this relation, we measure the viscosity $\eta$ from the Poiseuille flow profiles and fit a linear equation to the data. The viscosity is found using
\begin{equation}
	\eta = \frac{\rho a_y H^2}{8 u_{y}^{max}}.
\end{equation}

Fig. \ref{fig:Poiseuille}(c) shows the measured $\eta$ as a function of temperature for both relaxation times. As can be seen, the viscosity versus temperature data fits perfectly to a line and the slope of the lines correspond to the correct value of $\rho \tau$ for both systems confirming Eq. (\ref{eq:visc-T_relation}).

\subsection{Fourier Analysis}
The dynamic/shear viscosity and bulk viscosity of a fluid are related to the rate of decay of density waves in the fluid. Thus, we initially introduce a sound wave in the system and measure the shear and bulk viscosity from the decay of such waves.  We first compute what we expect theoretically for the linearized equations (small amplitude sound waves).
The continuity and momentum equations are
\begin{align}
	\partial_t \rho + \partial_{\alpha} (\rho u_{\alpha}) &= 0 \label{eq:FA_continuity}\\
	\partial_t (\rho u_{\beta}) + \partial_{\alpha}(\rho u_{\alpha} u_{\beta}) &= - \partial_{\beta}(\rho T) + \partial_{\alpha} \bigg(\eta (\partial_{\beta} u_{\alpha} + \partial_{\alpha} u_{\beta}) \bigg). \label{eq:FA_momentum}
\end{align}

We introduce a small perturbation in the form of a right traveling wave with the wave number $k$ and angular frequency $\omega$ ( $k, \omega>0$ )
\begin{align}
	&\rho=\rho_0 + a_0 e^{-\gamma t} e^{i(k x-\omega t)}\label{eq:rho_perturb}\\
	& u=b_0 e^{-\gamma t} e^{i(k x-\omega t)} \label{eq:velo_perturb}
\end{align}
where $a_0, \text{ and } b_0$ are constants with no dependence on space or time. The constants may be complex but they are small in value:
\begin{equation}
	a_0=\left|a_0\right| e^{i \delta_a}, \qquad \left|a_0\right| \sim 0
\end{equation}

If we plug (\ref{eq:rho_perturb}) and (\ref{eq:velo_perturb}) into (\ref{eq:FA_continuity}), we get
\begin{equation}
a_0(-\gamma-i \omega) e^{-\gamma t} e^{i(k x-\omega t)}+\rho_0 b_0(i k) e^{-\gamma t} e^{i(k x-\omega t)}+\mathcal{O}\left(a_0 b_0\right)=0
\end{equation}
which in turn gives
\begin{equation}\label{eq:FA_b0}
	b_0=a_0 \frac{(\gamma+i \omega)}{i k \rho_0}.
\end{equation}
The complex coefficient relating $a_0$ and $b_0$ implies $\rho$ and $u$ are not in phase but have a phase shift. Similarly, if we plug (\ref{eq:rho_perturb})-(\ref{eq:velo_perturb}) into the momentum equation (\ref{eq:FA_momentum}), we get
\begin{equation}
	 (\gamma + i \omega)^2 = -k^2 \left(T \right) + 2 k^2 \frac{\eta }{\rho_0} (\gamma+i \omega).
\end{equation}
An equation for the decay rate $\gamma$ can be found by equating the imaginary parts on each side of this equation: 
\begin{equation}\label{eq:FA_gamma}
	\gamma=k^2 \frac{\eta}{\rho_0}
\end{equation}

And the real part can be solved to get a dispersion relation:
\begin{equation}\label{eq:FA_dispersion}
\gamma^2-\omega^2=-k^2 T + 2 \frac{\eta}{\rho_0} k^2 \gamma.
\end{equation}
Substituting Eq. (\ref{eq:FA_gamma}) in Eq. (\ref{eq:FA_dispersion}), we get an equation for $\omega$ 
\begin{equation}\label{eq:FA_omega}
	\omega^2=k^2 T - \gamma^2
\end{equation}

Eq. (\ref{eq:FA_gamma}) and (\ref{eq:FA_omega}) can be used to measure $\eta$ from a sound wave analysis. To do so, we initialize our system with a sinusoidal perturbation
\begin{equation}
	\rho_{ijk} = \rho_0 + A cos(k z)
\end{equation}
where $\rho_0$ is the base density, $A$ is the amplitude of the wave, and $k=\frac{2\pi}{\lambda}$ is the wave number. For all the simulations, $\rho_0 = 1$ and $A=0.0001$ unless mentioned otherwise.

\begin{figure}[!htpb]
	\centering
	\begin{subfigure}[t]{0.492\textwidth}
		\includegraphics[width=\textwidth]{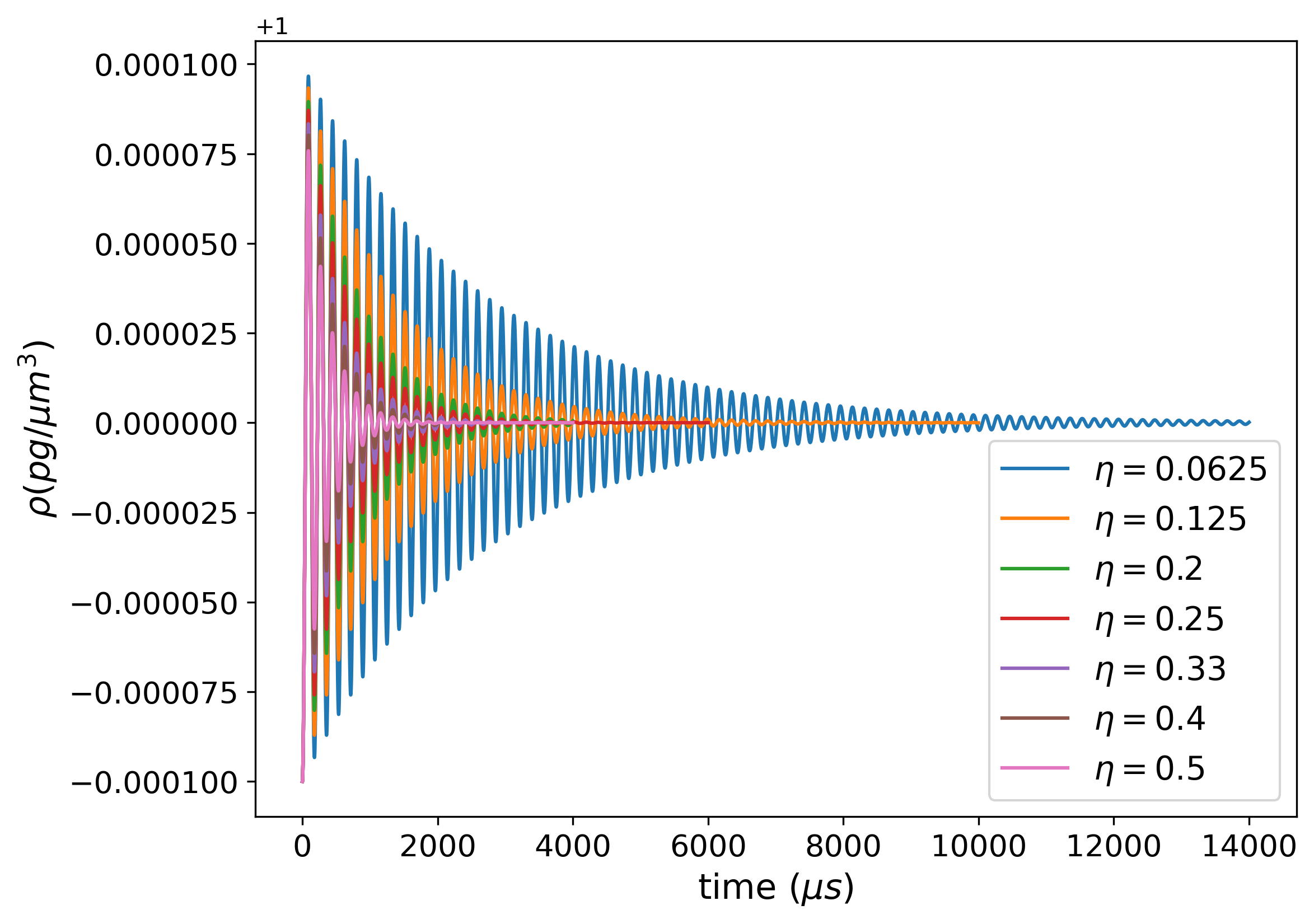}
		\caption{}
	\end{subfigure}
	\begin{subfigure}[t]{0.492\textwidth}
		\includegraphics[width=\textwidth]{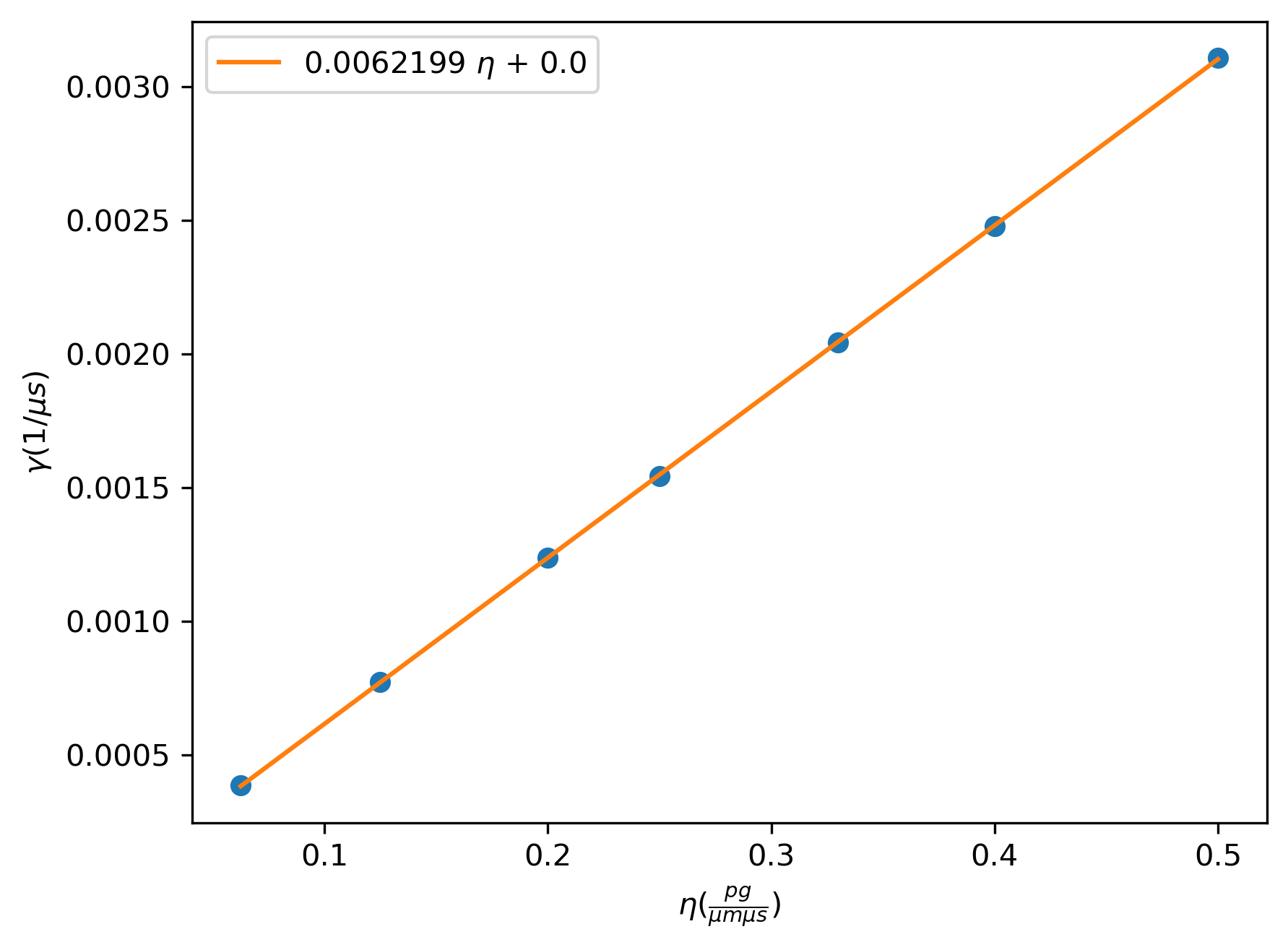}
		\caption{}
	\end{subfigure}
	\caption{\footnotesize{(a) shows the density oscillation at z=0 for different viscosities as a function of time. (b) shows the damping factor of the sound waves as function of viscosity. Higher viscosities damp out the waves faster as expected. A linear relation between $\gamma$ and $\eta$ is found as predicted by the wave analysis. The slope of the line is $k^2/\rho_0$}}
	\label{fig:soundwave_diffvis}
\end{figure}

We first look at the effect of viscosity on the decay of the sound wave. For these simulations, the simulation box dimensions are $20 \mu \text{m} \times 20 \mu \text{m} \times 80 \mu \text{m}$ and $k = 2\pi/80 = 0.07854$.
Fig. \ref{fig:soundwave_diffvis}(a) shows the evolution of the density at the centre of the simulation box. The results for viscosities $\eta = 0.0625, 0.125, 0.2, 0.25, 0.33, 0.4, 0.5$ are shown. We see that the waves are damped out faster at higher viscosities. This qualitative observation agrees with the prediction of Eq. (\ref{eq:FA_gamma}). To check if our model reproduces the correct effective viscosity, we fit the $\rho$ versus $t$ data to find the damping factor $\gamma$. Fig. \ref{fig:soundwave_diffvis} shows the measured damping factor as a function of the input viscosity. As can be seen, the damping factor is a linear function of the viscosity. The slope of the line is found to be $0.00622$ which matches the expected value $k^2/\rho_0 = 0.00617$ to within $1 \%$ error.
Moreover, one would expect changing the effective viscosity $\eta$ to have a negligible impact on the angular frequency as it only appears as the coefficient of $k^4$ in the $\gamma^2$ term. This is also observed in Fig. \ref{fig:soundwave_diffvis}(a) where the frequency of the waves do not differ substantially.

\begin{figure}
	\centering
	\begin{subfigure}[t]{0.492\textwidth}
		\includegraphics[width=\textwidth]{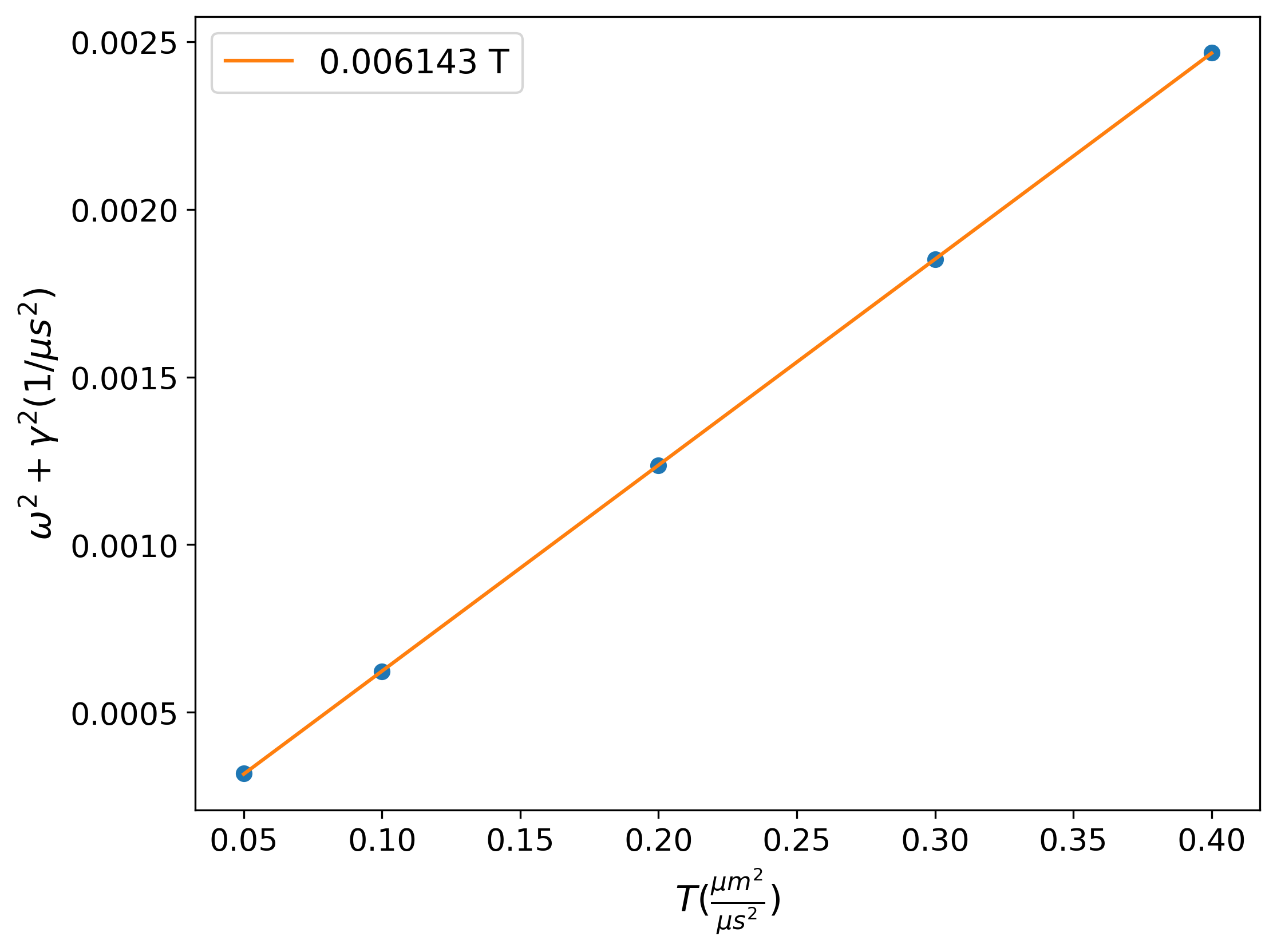}
		\caption{}
	\end{subfigure}
    	\begin{subfigure}[t]{0.492\textwidth}
		\includegraphics[width=\textwidth]{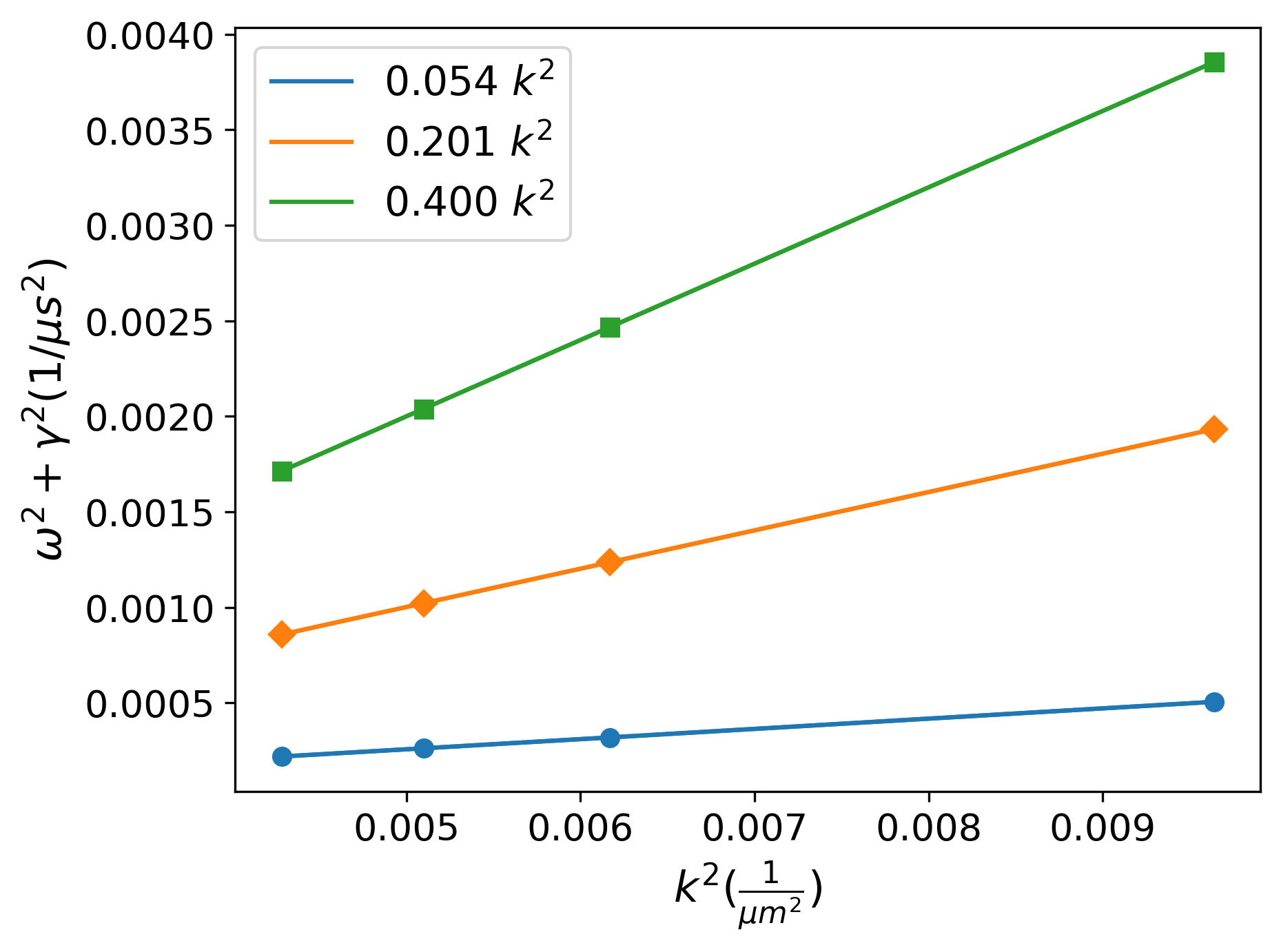}
		\caption{}
	\end{subfigure}
	\caption{\footnotesize (a) The sum of squared angular frequency and squared decay factor $\omega^2 + \gamma^2$ is plotted versus $T$ for $\eta=0.125$. The fitted line has a slope of $0.006153 \frac{1}{\mu \text{m}^2}$ which agrees with $k^2 \approx 0.006169$ within margin of error. (b) $\omega^2 + \gamma^2$ as a function of $k^2$ is shown for $T= 0.05, 0.2, 0.4\frac{\mu \text{m}^2}{\mu \text{s}^2}$. The slope of the lines is the same as temperature as expected from Eq. \ref{eq:FA_omega}}
	\label{fig:omega_diffT}
\end{figure}

One of the advantages of our new LBM is the flexibility in setting the temperature. Fig. \ref{fig:omega_diffT}(a) shows the square of the measured angular frequency as a function of temperature for input viscosity $\eta = 0.125 \frac{\text{pg}}{\mu \text{m} \mu \text{s}}$. A linear relationship between the $\omega^2+\gamma^2$ and $T$ is observed. The slope of the fitted lines correctly correspond to the value of $k^2$ with less than 1\% error.

We also look at the relation between the wave number $k$ and the dispersion relation. Fig. \ref{fig:omega_diffT}(b) shows $\omega^2+\gamma^2$ as a function of $k^2$ for three different temperatures. As can be seen, a linear relation between $\omega^2+\gamma^2$ and $k^2$ is recovered with a slope equal to the temperature.

So far we have shown that the fluid properties such as shear and bulk viscosities match the predicted values from the Chapman-Enskog expansion confirming that our model indeed recovers the correct Navier-Stokes equations without the presence of the error terms discussed in the introduction. One of the main consequences of eliminating such error terms is restoring Galilean invariance which will be discussed in the next section.

\subsection{Couette Flow and Galilean Invariance}

As discussed in section \ref{sec:intro}, the standard LBM on standard lattices is not capable of generating independent third moments. This leads to appearance of error terms in the macroscopic momentum equation and limits the applicability of the model to the incompressible limit. The error terms also introduce a dependence on the frame and break the Galilean invariance of the model \cite{Kruger2016book, Holdych1998}. These error terms are typically negligible in the incompressible and low velocity regimes but significant when a density gradient is applied. In this section, we show that our model is Galilean invariant even in the presence of significant density gradients in the compressible regime.

\begin{figure}[!bthp]
 	\centering
 	\includegraphics[width=0.7\textwidth]{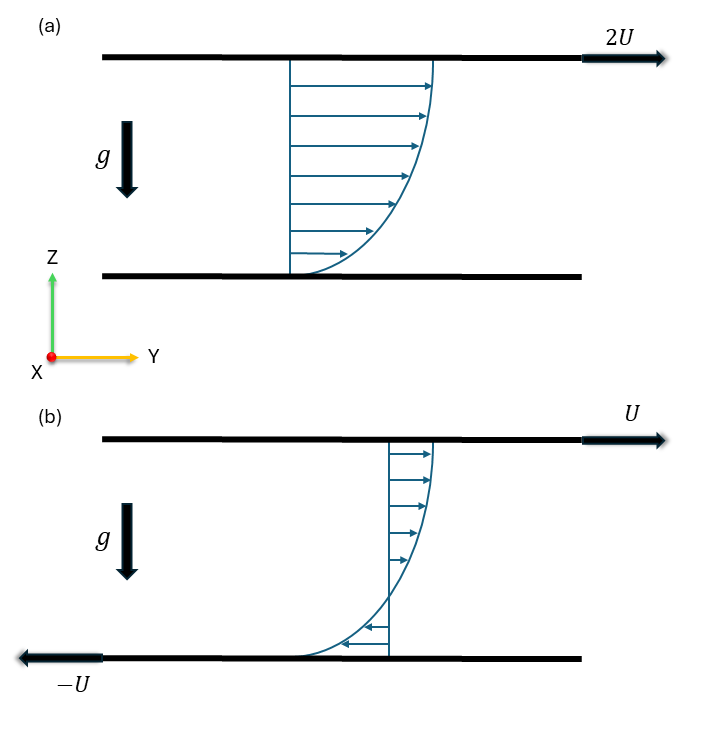}
 	\caption{\footnotesize{The velocity profile for a Couette flow in the presence of a gravitational field is depicted. (a) stationary bottom wall and moving top wall (b) top and bottom walls moving at the same velocity in the opposite directions}}
 	\label{fig:Couette_cartoon}
 \end{figure}

We test the Galilean invariance of our system in a Couette flow example. Two simulations are conducted: In the first system, the bottom wall is stationary and the top wall moves in the y direction at the speed of $2U$. In the second system, the bottom wall moves at the speed of $U$ in the -y direction and the top wall moves in the y direction at the speed of $U$. In both systems, a body force is applied in the $-z$ direction to create a density gradient. A periodic boundary condition is applied in the $x$ and $y$ directions. The solid walls are set up normal to the $z$ direction and bounce back boundary conditions are invoked. Schematic diagrams of both systems are illustrated in Fig. \ref{fig:Couette_cartoon}.

\begin{figure}
 	\centering
 	\includegraphics[width=0.7\textwidth]{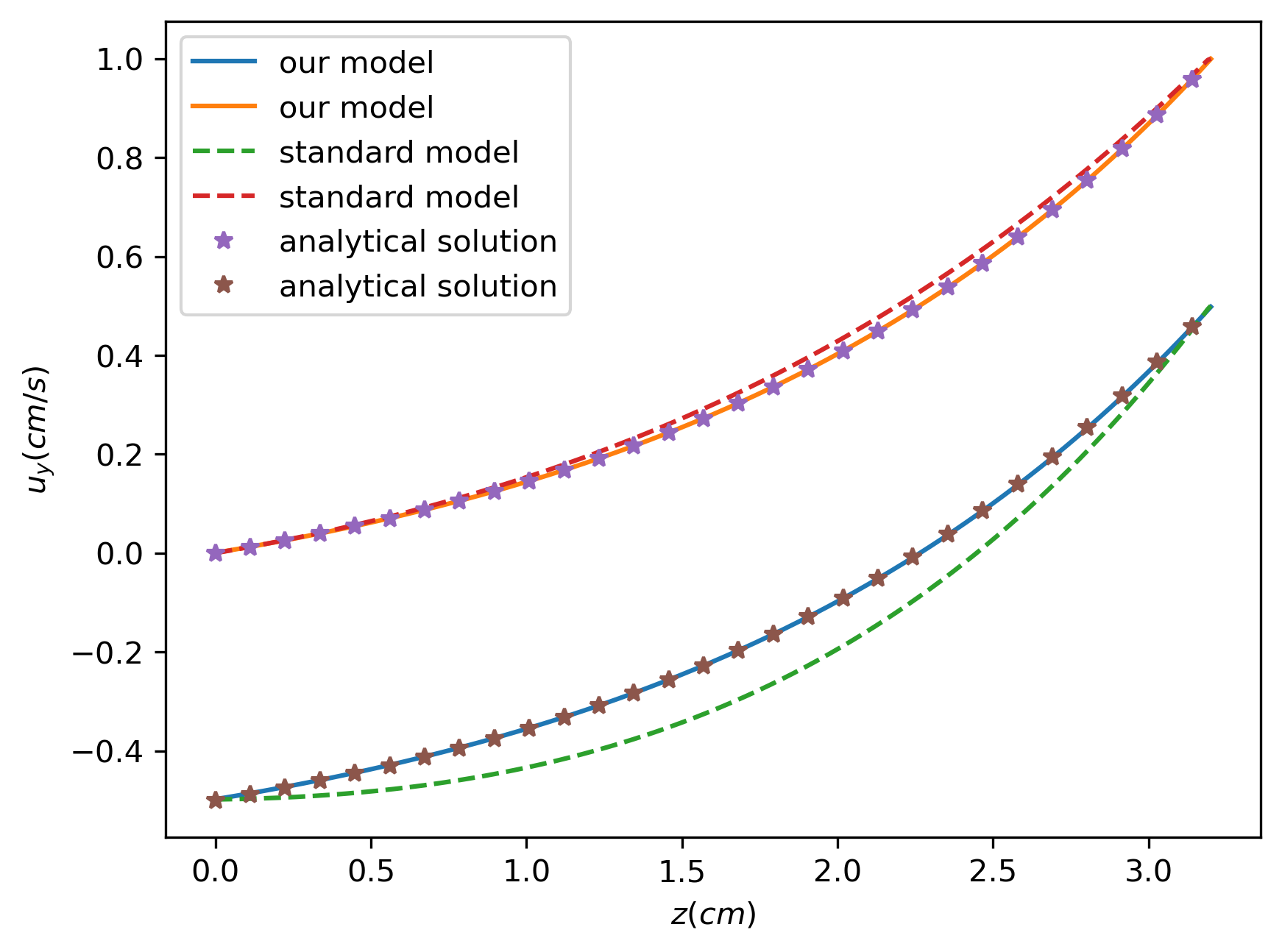}
 	\caption{\footnotesize{The velocity profile obtained from standard LB, our model, and analytical solutions for two types of Couette flow are shown. Our model produces velocity profiles that exactly match the analytical solution (star points) and do not depend on the frame of reference.}}
 	\label{fig:Couette}
 \end{figure}

The velocity profile for this problem can be analytically found by solving Eq. (\ref{eq:LB_NSE}): 
\begin{equation}\label{eq:couette_momentum}
	\partial_{t} (\rho u_{\beta}) + \partial_{\alpha}(\rho u_{\alpha} u_{\beta}) = -\partial_{\beta}(\rho T) + \partial_{\alpha}\bigg(\eta(\partial_\alpha u_\beta+\partial_\beta u_\alpha) \bigg) + \rho a_\beta
\end{equation}
where $\rho a_\beta$ is the body force term. For the above geometry and under steady state conditions, Eq. (\ref{eq:couette_momentum}) can be simplified to get
\begin{align}
	0 &=  \tau T \partial_z \bigg(\rho(\partial_z u_y) \bigg), \qquad u_y (z=0) = u_b, u_y (z=H) = u_t, \label{eq:couette_y}\\
	0 &=  - T \partial_z \rho + \rho a_z. \label{eq:couette_z}
\end{align}
where $H$ is the hight of the channel (distance between the plates), $u_b$ is the velocity of the bottom plate, and $u_t$ is the velocity of the top wall. The second equation produces an exponential density profile as a function of $z$, and by solving the system of equations we find the velocity profile
\begin{equation}
	u_y = (u_t - u_b)\bigg[\frac{1-e^{-(a_z/T)z}}{1-e^{-(a_z/T)H}}\bigg] + u_b.
\end{equation}
In our system, $a_z = -g = -980 \text{ cm}/\text{s}^2$, $U = 0.5 \text{ cm}/\text{s}$, $T = 1/6 \text{ cm}^2/\text{s}^2$, and $\rho = 0.001184 \text{ g}/\text{cm}^3$. Fig. \ref{fig:Couette} demonstrates the velocity profiles obtained from analytical solution above as well as the results for the standard LB and our new model. The profiles for the current model, the standard model, and the analytical solution are shown as solid lines, dashed lines, and star points. As can be seen, switching the frame of reference results in a new profile in the standard model and Galilean invariance is broken. On the other hand, the current model not only respects Galilean invariance but also exactly matches the analytical solution (i.e. the profile is exactly the same for the two cases other than the shift by $U$).

\subsection{Flow over Cylinder}
One of the advantages of the new method is that it decouples the mesh velocity and the speed of sound. In this section, we test the compressibility effects as well as the impact of temporal and spatial resolution on the accuracy of the LB simulations for a flow over a cylinder.

The simulations are performed in 3D and box dimensions are 0.06 cm $\times$ 2.4 cm $\times$ 4.8 cm. Periodic boundary conditions are applied in the x and y directions. Fixed walls in the z direction are applied using the bounce back method \cite{Ladd1994}. The density of the fluid is set to that of air (0.001184 $\text{g}/\text{cm}^3$) and the viscosity is about 2.5 times of that of air (0.05 cP). The speed of sound is fixed at $50 \text{cm}/{s}$. An acceleration of $10^{-5} \text{cm}/\text{s}^2$ is applied in the y-direction. The main axis of the cylinder is in the x direction. The diameter and height of the cylinder are 0.2 cm and 0.06 cm, respectively. The fluid-structure interactions are implemented as described in \cite{Denniston2022}.

\begin{figure}
	\centering
	\includegraphics[width=\textwidth]{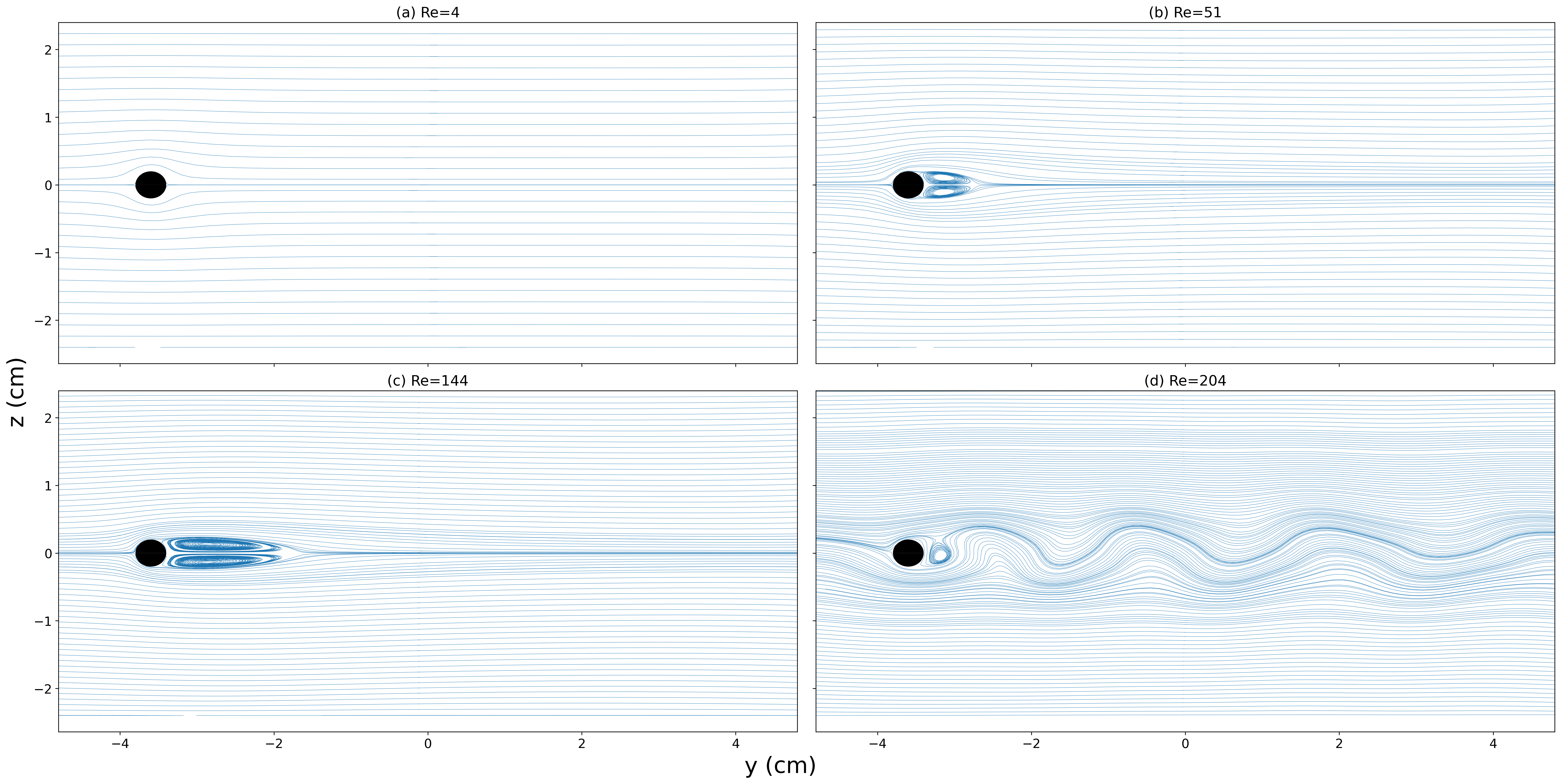}
	\caption{\footnotesize Flow streamlines are shown for (a) $Re$=4 (b) $Re$=51 (c) $Re$=144 (d) $Re$=204}
	\label{fig:streamlines}
\end{figure}

The flow is started from rest and gradually increases with time allowing us to effectively sample different Reynolds numbers as a function of time.  The velocity streamlines are depicted in Fig. \ref{fig:streamlines} at four different Reynolds numbers for a system with lattice spacing $\Delta x = 0.01 $ cm and time step $\Delta t = 0.0001$ s. Reynolds number is defined as
\begin{equation}
	Re = \frac{\rho u_\infty D}{\eta}
\end{equation}
where $\rho$ is the density, $u_\infty$ is the far-stream velocity, $D$ is the cylinder diameter, and $\eta$ is the dynamic/shear viscosity. Different flow regimes are observed depending on Reynolds number. Initially, the velocity and consequently $Re$ are small. Therefore, we see a creeping flow over the cylinder in Fig. \ref{fig:streamlines}(a). As the velocity increases, we see separation of the fluid streamlines and formation of vortices behind the cylinder. The size of the vortices grow with the Reynolds number $Re$. Fig. \ref{fig:streamlines}(c) shows the streamlines for $Re = 144$ and the separation vortices at their maximum size before onset of vortex shedding. The maximum length of the separation vortices is 1.4 cm. For $Re > 190$, we observe vortex shedding and turbulent behaviour.

\begin{figure}
	\centering
	\includegraphics[width=\textwidth]{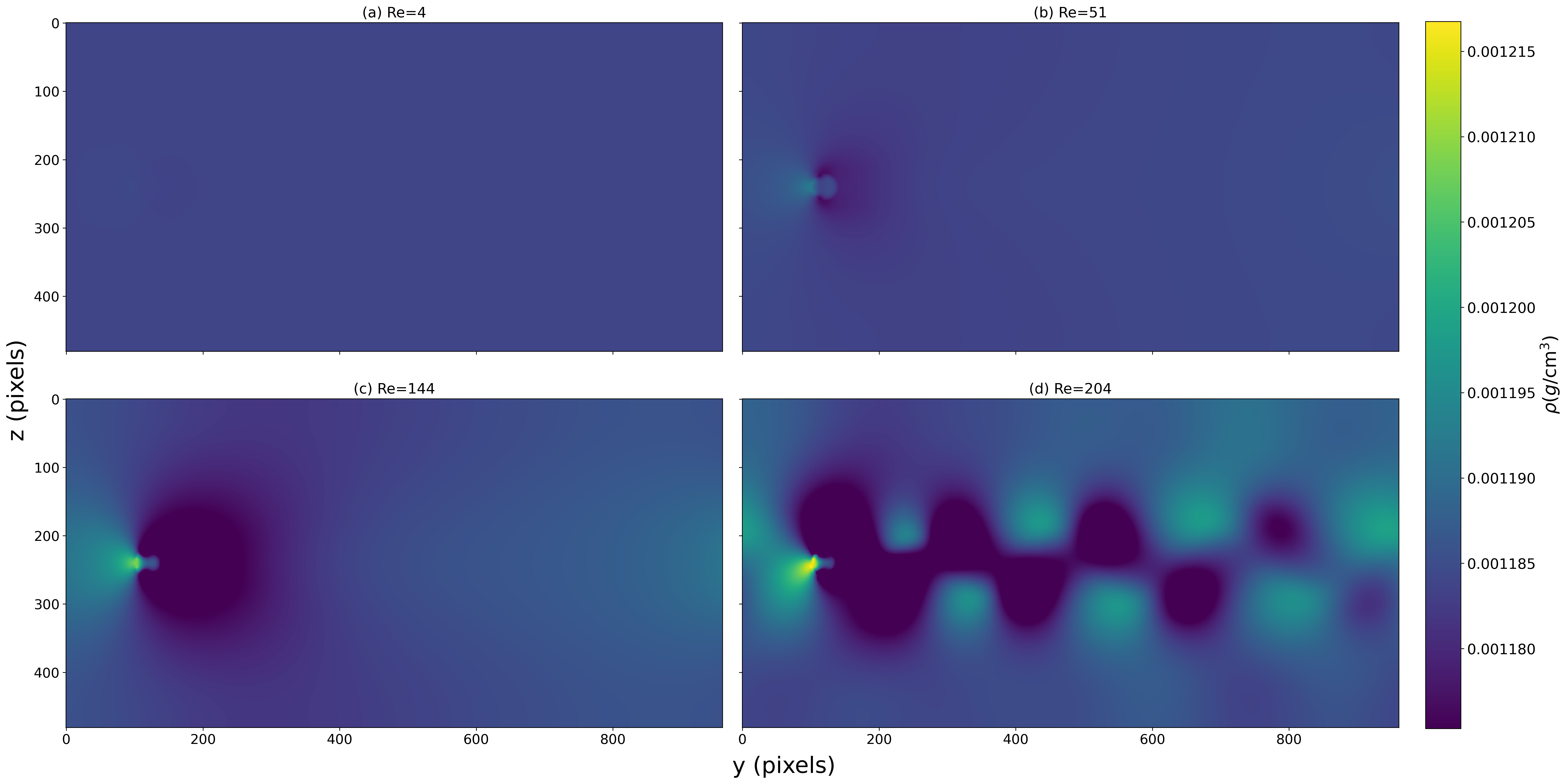}
	\caption{\footnotesize Density contours are shown at different Reynolds numbers (a) $Re$=4 (b) $Re$=51 (c) $Re$=144 (d) $Re$=204. For $Re > 190$, vortex shedding and considerable density gradients are observed}
	\label{fig:rho_heatmap}
\end{figure}

In the turbulent regime, the compressibility effects become considerable. Fig. \ref{fig:rho_heatmap} shows the density contours at the 4 different Reynolds numbers. As can be seen, the density is uniform at very low velocities. With the increase of the velocity, we observe formation of high pressure points in front of the cylinder. This is shown as green regions in Fig. \ref{fig:rho_heatmap}(b) and (c). After onset of vortex shedding, we see regions of high and low density intermittently creating a wake behind the object as demonstrated in Fig. \ref{fig:rho_heatmap}(d). At $Re = 204$,  the variations in density are 8\% and Mach number is 0.5.

\begin{figure}
	\centering
	\includegraphics[width=0.5\textwidth]{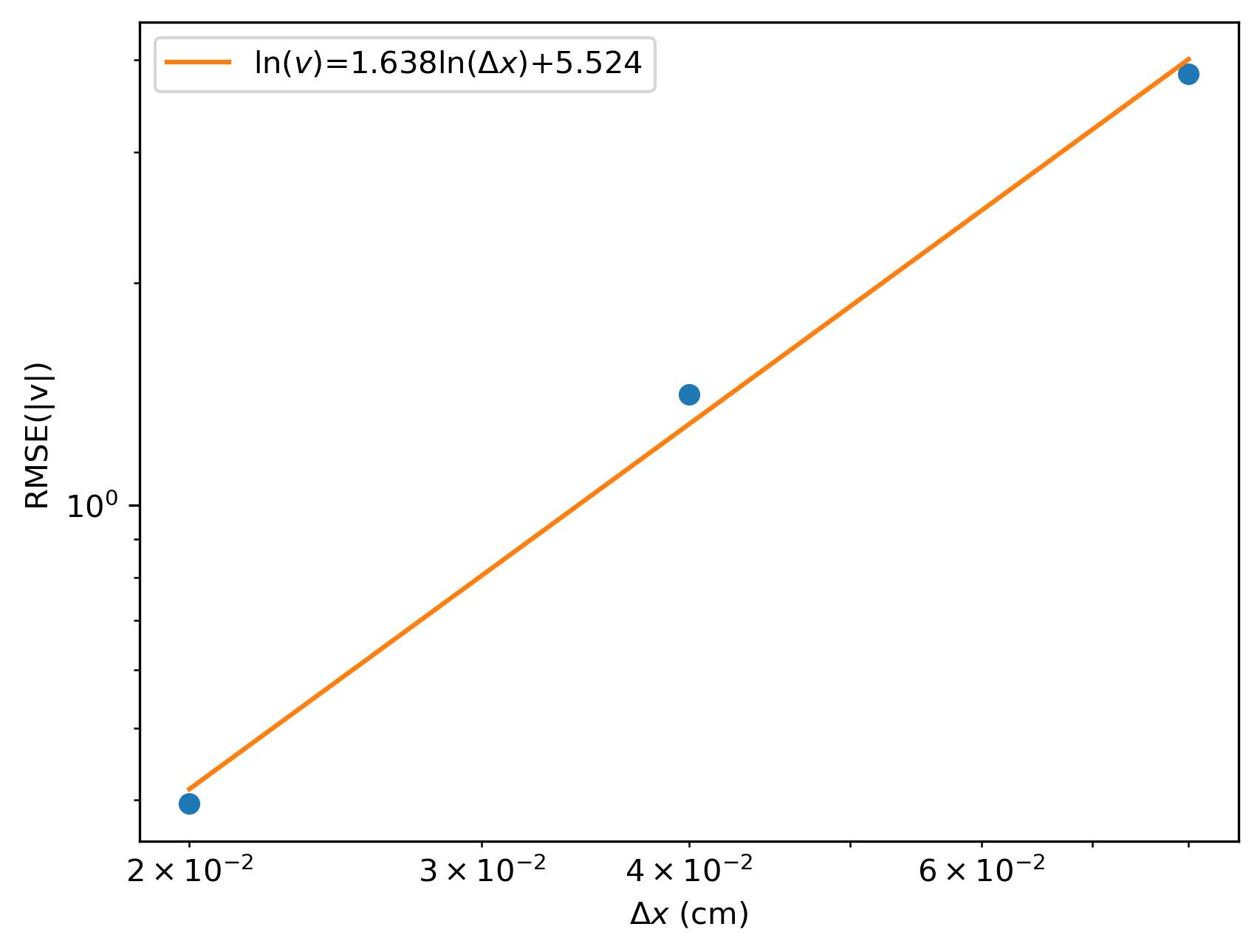}
	\caption{\footnotesize The root-mean-square error is shown as a function of lattice spacing}
	\label{fig:rmse}
\end{figure}

To see the impact of mesh resolution on the accuracy of the model, we compared 4 simulations with different lattice spacings. For these simulations, $\Delta t =0.0001$ s. Lattice spacings of $\Delta x = 0.01, 0.02, 0.04, 0.08$ cm are compared with the smallest being set as the ground truth. We compare the simulations in the laminar regime. Fig. \ref{fig:rmse}(a) shows the root-mean-square error on a logarithmic scale. The fitted line has a slope of 1.638. This is close to the second order spatial discretization error expected from Lattice Boltmzann. We expect the fluid-structure errors to be the main reason for this slightly below second-order relation.

\section{Conclusion}\label{sec:conclusion}
The Lattice Botlzmann Method (LBM) has been a powerful tool for modeling fluid dynamics. The standard LBM on a standard lattice (i.e. $e_{i\alpha}=\Delta x$) results in error terms of form $u_\beta \partial_\alpha \rho$ and $\partial_\gamma(\rho u_\alpha u_\beta u_\gamma)$ in the macroscopic equation. This is typically addressed by applying the equation of state $P_{\alpha \beta}=\rho \frac{c^3}{3}\delta_{\alpha \beta}$. However, it results in a coupling between the equation of state and the lattice velocity limiting the application of LBM. In this paper, we introduce a new velocity discretization method that results in the correct macroscopic equations of fluid dynamics on standard lattices and extends the application of the model to compressible flows.

We show that the standard velocity discretization is equivalent to an expansion of the particle population function $f(\bm{x}, \bm{v}, t)$ using delta functions:
\begin{equation}
    f(\bm{x}, \bm{v}, t) = \sum_i f_i \delta(v-e_i).
\end{equation}
Leveraging this new interpretation, we substitute the delta functions with bump functions:
\begin{equation}
    f(\bm{x}, \bm{v}, t) = \sum_i f_i p(e_i, b_{\alpha \beta})
\end{equation}
where $b_{\alpha \beta}$ terms are related to the variance/width of the bump functions. Introducing the $b$ terms provides enough degrees of freedom to set the equilibrium third moments of the discrete distribution to that of the Maxwell-Boltzmann distribution. This leads to recovery of the correct macroscopic equations without the error terms.

The proposed method is evaluated for both incompressible and compressible flow regimes. In the case of body force-driven flow between two stationary parallel plates, the method accurately reproduces the Poiseuille velocity profile and the corresponding shear viscosity.  
The attenuation of sound waves in a fluid is also examined, with the method successfully capturing the expected decay coefficient and dispersion relation.  
For Couette flow under the influence of gravity, the method yields a velocity profile in excellent agreement with the analytical solution while maintaining Galilean invariance.
Finally, we studied the impact of mesh resolution on the accuracy of the simulations. 

\section{Data Availability Statement}
The data that supports the findings of this study are available within the article.

\begin{acknowledgments}
This work was supported by the Natural Science and Engineering Council of Canada (NSERC) Discovery grant (CD) and a CREATE grant (NA) on Advanced Polymer Composite Materials and Technologies.  Computational resources were provided by the Shared Hierarchical Academic Research Computing Network (SHARCNET) and the Digital Research Alliance of Canada.  CD acknowledges useful discussions with STT Ollila on some of the initial ideas for this work. 
\end{acknowledgments}
\appendix
\section{Equilibrium distributions}\label{app:eq_dist}
The equilibrium particle distributions $f^{eq}_i$ and equilibrium $b_{\alpha \beta}$ are found by matching the moments of continuous $f$ with those of Maxwell-Boltzmann distribution. Solving the system of linear equations for a $D3Q19$ model, we find the equilibrium values to be
\small
\begin{align*}
f^{eq}_0 &= \frac{\rho  \left(3 \text{c}^4-2 \left(u_x^2+u_y^2+u_z^2\right) \text{c}^2-6 \left(\left(u_y^2+u_z^2\right) u_x^2+u_y^2 u_z^2\right)\right)}{9 \text{c}^4} \\
f^{eq}_1 &= \frac{\rho  \left(\text{c}^4+3 u_x \text{c}^3+2 \left(u_x^2-u_y^2-u_z^2\right) \text{c}^2+6 u_x^2 \left(u_y^2+u_z^2\right)\right)}{18 \text{c}^4} \\
f^{eq}_2 &= \frac{\rho  \left(\text{c}^4+3 u_y \text{c}^3-2 \left(u_x^2-u_y^2+u_z^2\right) \text{c}^2+6 u_y^2 \left(u_x^2+u_z^2\right)\right)}{18 \text{c}^4} \\
f^{eq}_3 &= \frac{\rho  \left(\text{c}^4-3 u_x \text{c}^3+2 \left(u_x^2-u_y^2-u_z^2\right) \text{c}^2+6 u_x^2 \left(u_y^2+u_z^2\right)\right)}{18 \text{c}^4} \\
f^{eq}_4 &= \frac{\rho  \left(\text{c}^4-3 u_y \text{c}^3-2 \left(u_x^2-u_y^2+u_z^2\right) \text{c}^2+6 u_y^2 \left(u_x^2+u_z^2\right)\right)}{18 \text{c}^4} \\
f^{eq}_5 &= \frac{\rho  \left(\text{c}^4+3 u_z \text{c}^3-2 \left(u_x^2+u_y^2-u_z^2\right) \text{c}^2+6 \left(u_x^2+u_y^2\right) u_z^2\right)}{18 \text{c}^4} \\
f^{eq}_6 &= \frac{\rho  \left(\text{c}^4-3 u_z \text{c}^3-2 \left(u_x^2+u_y^2-u_z^2\right) \text{c}^2+6 \left(u_x^2+u_y^2\right) u_z^2\right)}{18 \text{c}^4} \\
f^{eq}_7 &= \frac{\rho  \left(\text{c}^4+3 (u_x+u_y) \text{c}^3+2 \left(u_x^2+3 u_y u_x+u_y^2\right) \text{c}^2-6 u_x^2 u_y^2\right)}{36 \text{c}^4} \\
f^{eq}_8 &= \frac{\rho  \left(\text{c}^4+3 (u_x-u_y) \text{c}^3+2 \left(u_x^2-3 u_y u_x+u_y^2\right) \text{c}^2-6 u_x^2 u_y^2\right)}{36 \text{c}^4} \\
f^{eq}_9 &= \frac{\rho  \left(\text{c}^4-3 (u_x-u_y) \text{c}^3+2 \left(u_x^2-3 u_y u_x+u_y^2\right) \text{c}^2-6 u_x^2 u_y^2\right)}{36 \text{c}^4} \\
f^{eq}_{10} &= \frac{\rho  \left(\text{c}^4-3 (u_x+u_y) \text{c}^3+2 \left(u_x^2+3 u_y u_x+u_y^2\right) \text{c}^2-6 u_x^2 u_y^2\right)}{36 \text{c}^4} \\
	f^{eq}_{11} &= \frac{\rho  \left(\text{c}^4+3 (u_x+u_z) \text{c}^3+2 \left(u_x^2+3 u_z u_x+u_z^2\right) \text{c}^2-6 u_x^2 u_z^2\right)}{36 \text{c}^4} \\
f^{eq}_{12} &= \frac{\rho  \left(\text{c}^4+3 (u_x-u_z) \text{c}^3+2 \left(u_x^2-3 u_z u_x+u_z^2\right) \text{c}^2-6 u_x^2 u_z^2\right)}{36 \text{c}^4} \\
f^{eq}_{13} &= \frac{\rho  \left(\text{c}^4-3 (u_x-u_z) \text{c}^3+2 \left(u_x^2-3 u_z u_x+u_z^2\right) \text{c}^2-6 u_x^2 u_z^2\right)}{36 \text{c}^4} \\
f^{eq}_{14} &= \frac{\rho  \left(\text{c}^4-3 (u_x+u_z) \text{c}^3+2 \left(u_x^2+3 u_z u_x+u_z^2\right) \text{c}^2-6 u_x^2 u_z^2\right)}{36 \text{c}^4} \\
f^{eq}_{15} &= \frac{\rho  \left(\text{c}^4+3 (u_y+u_z) \text{c}^3+2 \left(u_y^2+3 u_z u_y+u_z^2\right) \text{c}^2-6 u_y^2 u_z^2\right)}{36 \text{c}^4} \\
f^{eq}_{16} &= \frac{\rho  \left(\text{c}^4+3 (u_y-u_z) \text{c}^3+2 \left(u_y^2-3 u_z u_y+u_z^2\right) \text{c}^2-6 u_y^2 u_z^2\right)}{36 \text{c}^4} \\
f^{eq}_{17} &= \frac{\rho  \left(\text{c}^4-3 (u_y-u_z) \text{c}^3+2 \left(u_y^2-3 u_z u_y+u_z^2\right) \text{c}^2-6 u_y^2 u_z^2\right)}{36 \text{c}^4} \\
f^{eq}_{18} &= \frac{\rho  \left(\text{c}^4-3 (u_y+u_z) \text{c}^3+2 \left(u_y^2+3 u_z u_y+u_z^2\right) \text{c}^2-6 u_y^2 u_z^2\right)}{36 \text{c}^4}
\end{align*}

\small
\begin{align*}
	b_{xx}^{eq} &= \frac{1}{3} \left(-\text{c}^2+u_x^2+3 T \right),\, b_{yy}^{eq}= \frac{1}{3} \left(-\text{c}^2+u_y^2+3 T \right),\,
	b_{zz}^{eq} = \frac{1}{3} \left(-\text{c}^2+u_z^2+3 T \right), \\
	b_{xy}^{eq} &= \frac{1}{3} u_x u_y,\, b_{xz}^{eq}= \frac{1}{3} u_x u_z, \,b_{yz}^{eq}= \frac{1}{3} u_y u_z
\end{align*}
\section{Transformation matrix}\label{app:trans_matrix}
The transformation matrix between the particle distribution space and the moment space is found from matching moments with the moments of the Maxwell-Boltzmann distribution and solving for the equilibrium particle distribution function:
\begin{equation}
    f^{eq}_i (\bm{x}, t) = w_i \sum_i m^a_i M^a_{eq}(\bm{x}, t) N^a 
\end{equation}
where $w_i$ are the weight factors, $\bm{m}$ is the transformation matrix from population to moment space, $M^a_{eq}(\bm{x}, t)$ are the equilibrium moments, and $N^a$ are the normalization constants. The transformation matrix is presented in Table \ref{table:M}. The particle distribution function can be computed using the same transformation:
\begin{equation}
    f_i (\bm{x}, t) = w_i \sum_i m^a_i M^a(\bm{x}, t) N^a 
\end{equation}
{\footnotesize
\begin{table}
	\centering
	\begin{equation*}
		\setlength\arraycolsep{0.5pt}
		\left(
		\begin{array}{ccccccccccccccccccc}
			1 & 1 & 1 & 1 & 1 & 1 & 1 & 1 & 1 & 1 & 1 & 1 & 1 & 1 & 1 & 1 & 1 & 1 & 1 \\
			0 & \text{c} & 0 & -\text{c} & 0 & 0 & 0 & \text{c} & \text{c} & -\text{c} & -\text{c} & \text{c} & \text{c} & -\text{c} & -\text{c} & 0 & 0 & 0 & 0 \\
			0 & 0 & \text{c} & 0 & -\text{c} & 0 & 0 & \text{c} & -\text{c} & \text{c} & -\text{c} & 0 & 0 & 0 & 0 & \text{c} & \text{c} & -\text{c} & -\text{c} \\
			0 & 0 & 0 & 0 & 0 & \text{c} & -\text{c} & 0 & 0 & 0 & 0 & \text{c} & -\text{c} & \text{c} & -\text{c} & \text{c} & -\text{c} & \text{c} & -\text{c} \\
			-\frac{\text{c}^2}{3} & \frac{2 \text{c}^2}{3} & -\frac{\text{c}^2}{3} & \frac{2 \text{c}^2}{3} & -\frac{\text{c}^2}{3} & -\frac{\text{c}^2}{3} & -\frac{\text{c}^2}{3} & \frac{2 \text{c}^2}{3} & \frac{2 \text{c}^2}{3} & \frac{2 \text{c}^2}{3} & \frac{2 \text{c}^2}{3} & \frac{2 \text{c}^2}{3} & \frac{2 \text{c}^2}{3} & \frac{2 \text{c}^2}{3} & \frac{2 \text{c}^2}{3} & -\frac{\text{c}^2}{3} & -\frac{\text{c}^2}{3} & -\frac{\text{c}^2}{3} & -\frac{\text{c}^2}{3} \\
			-\frac{\text{c}^2}{3} & -\frac{\text{c}^2}{3} & \frac{2 \text{c}^2}{3} & -\frac{\text{c}^2}{3} & \frac{2 \text{c}^2}{3} & -\frac{\text{c}^2}{3} & -\frac{\text{c}^2}{3} & \frac{2 \text{c}^2}{3} & \frac{2 \text{c}^2}{3} & \frac{2 \text{c}^2}{3} & \frac{2 \text{c}^2}{3} & -\frac{\text{c}^2}{3} & -\frac{\text{c}^2}{3} & -\frac{\text{c}^2}{3} & -\frac{\text{c}^2}{3} & \frac{2 \text{c}^2}{3} & \frac{2 \text{c}^2}{3} & \frac{2 \text{c}^2}{3} & \frac{2 \text{c}^2}{3} \\
			-\frac{\text{c}^2}{3} & -\frac{\text{c}^2}{3} & -\frac{\text{c}^2}{3} & -\frac{\text{c}^2}{3} & -\frac{\text{c}^2}{3} & \frac{2 \text{c}^2}{3} & \frac{2 \text{c}^2}{3} & -\frac{\text{c}^2}{3} & -\frac{\text{c}^2}{3} & -\frac{\text{c}^2}{3} & -\frac{\text{c}^2}{3} & \frac{2 \text{c}^2}{3} & \frac{2 \text{c}^2}{3} & \frac{2 \text{c}^2}{3} & \frac{2 \text{c}^2}{3} & \frac{2 \text{c}^2}{3} & \frac{2 \text{c}^2}{3} & \frac{2 \text{c}^2}{3} & \frac{2 \text{c}^2}{3} \\
			0 & 0 & 0 & 0 & 0 & 0 & 0 & \text{c}^2 & -\text{c}^2 & -\text{c}^2 & \text{c}^2 & 0 & 0 & 0 & 0 & 0 & 0 & 0 & 0 \\
			0 & 0 & 0 & 0 & 0 & 0 & 0 & 0 & 0 & 0 & 0 & \text{c}^2 & -\text{c}^2 & -\text{c}^2 & \text{c}^2 & 0 & 0 & 0 & 0 \\
			0 & 0 & 0 & 0 & 0 & 0 & 0 & 0 & 0 & 0 & 0 & 0 & 0 & 0 & 0 & \text{c}^2 & -\text{c}^2 & -\text{c}^2 & \text{c}^2 \\
			0 & -\frac{2 \text{c}^3}{3} & 0 & \frac{2 \text{c}^3}{3} & 0 & 0 & 0 & \frac{\text{c}^3}{3} & \frac{\text{c}^3}{3} & -\frac{\text{c}^3}{3} & -\frac{\text{c}^3}{3} & \frac{\text{c}^3}{3} & \frac{\text{c}^3}{3} & -\frac{\text{c}^3}{3} & -\frac{\text{c}^3}{3} & 0 & 0 & 0 & 0 \\
			0 & 0 & -\frac{2 \text{c}^3}{3} & 0 & \frac{2 \text{c}^3}{3} & 0 & 0 & \frac{\text{c}^3}{3} & -\frac{\text{c}^3}{3} & \frac{\text{c}^3}{3} & -\frac{\text{c}^3}{3} & 0 & 0 & 0 & 0 & \frac{\text{c}^3}{3} & \frac{\text{c}^3}{3} & -\frac{\text{c}^3}{3} & -\frac{\text{c}^3}{3} \\
			0 & 0 & 0 & 0 & 0 & -\frac{2 \text{c}^3}{3} & \frac{2 \text{c}^3}{3} & 0 & 0 & 0 & 0 & \frac{\text{c}^3}{3} & -\frac{\text{c}^3}{3} & \frac{\text{c}^3}{3} & -\frac{\text{c}^3}{3} & \frac{\text{c}^3}{3} & -\frac{\text{c}^3}{3} & \frac{\text{c}^3}{3} & -\frac{\text{c}^3}{3} \\
			0 & 0 & 0 & 0 & 0 & 0 & 0 & \text{c}^3 & \text{c}^3 & -\text{c}^3 & -\text{c}^3 & -\text{c}^3 & -\text{c}^3 & \text{c}^3 & \text{c}^3 & 0 & 0 & 0 & 0 \\
			0 & 0 & 0 & 0 & 0 & 0 & 0 & \text{c}^3 & -\text{c}^3 & \text{c}^3 & -\text{c}^3 & 0 & 0 & 0 & 0 & -\text{c}^3 & -\text{c}^3 & \text{c}^3 & \text{c}^3 \\
			0 & 0 & 0 & 0 & 0 & 0 & 0 & 0 & 0 & 0 & 0 & \text{c}^3 & -\text{c}^3 & \text{c}^3 & -\text{c}^3 & -\text{c}^3 & \text{c}^3 & -\text{c}^3 & \text{c}^3 \\
			0 & \frac{\text{c}^4}{2} & \frac{\text{c}^4}{2} & \frac{\text{c}^4}{2} & \frac{\text{c}^4}{2} & -\text{c}^4 & -\text{c}^4 & -\text{c}^4 & -\text{c}^4 & -\text{c}^4 & -\text{c}^4 & \frac{\text{c}^4}{2} & \frac{\text{c}^4}{2} & \frac{\text{c}^4}{2} & \frac{\text{c}^4}{2} & \frac{\text{c}^4}{2} & \frac{\text{c}^4}{2} & \frac{\text{c}^4}{2} & \frac{\text{c}^4}{2} \\
			0 & \frac{\text{c}^4}{2} & -\frac{\text{c}^4}{2} & \frac{\text{c}^4}{2} & -\frac{\text{c}^4}{2} & 0 & 0 & 0 & 0 & 0 & 0 & -\frac{\text{c}^4}{2} & -\frac{\text{c}^4}{2} & -\frac{\text{c}^4}{2} & -\frac{\text{c}^4}{2} & \frac{\text{c}^4}{2} & \frac{\text{c}^4}{2} & \frac{\text{c}^4}{2} & \frac{\text{c}^4}{2} \\
			\frac{\text{c}^4}{6} & -\frac{\text{c}^4}{3} & -\frac{\text{c}^4}{3} & -\frac{\text{c}^4}{3} & -\frac{\text{c}^4}{3} & -\frac{\text{c}^4}{3} & -\frac{\text{c}^4}{3} & \frac{\text{c}^4}{6} & \frac{\text{c}^4}{6} & \frac{\text{c}^4}{6} & \frac{\text{c}^4}{6} & \frac{\text{c}^4}{6} & \frac{\text{c}^4}{6} & \frac{\text{c}^4}{6} & \frac{\text{c}^4}{6} & \frac{\text{c}^4}{6} & \frac{\text{c}^4}{6} & \frac{\text{c}^4}{6} & \frac{\text{c}^4}{6} \\
		\end{array}
		\right)
	\end{equation*}
	\caption{\footnotesize{The population-to-moment transformation. $\text{c} = \Delta x/\Delta t$ is the lattice velocity.}}
	\label{table:M}
\end{table}
}
\bibliographystyle{aipnum4-2.bst}
\bibliography{./compresslb}

\end{document}